\documentclass[isre]{informs3}
\usepackage[utf8]{inputenc}
\usepackage[english]{babel}
\usepackage{geometry}
\usepackage{setspace}
\usepackage{natbib}
 \bibpunct[, ]{(}{)}{,}{a}{}{,}%
 \def\bibfont{\small}%
 \renewcommand{\bibfont}{\fontsize{9}{10}\selectfont}
\usepackage{hyperref} % For hyperlinks
\usepackage{graphicx} % For figures
\usepackage{authblk} % For author affiliations
\usepackage{booktabs}
\usepackage{threeparttable}
\usepackage{amsmath}  % for the equation environment
\usepackage{tabularx}
\usepackage{multirow}
\usepackage{comment}

\DoubleSpacedXI
\def\FigureNoteStyle{\TABLEfootnotesizeIX\rm}
\TheoremsNumberedThrough  
\EquationsNumberedThrough 
\def\theARTICLETOP{}

\def\setoddRH{}
\def\setevenRH{}
\def\setoddRF{\raisebox{-24pt}[0pt][0pt]{\hbox to \textwidth{\hfil\fs.8.8.\thepage\hfil}}}
\def\setevenRF{\setoddRF}

\TITLE{The Differential Effects of AI Abstention on Advice Acceptance}
\ARTICLEAUTHORS{%
\AUTHOR{Nicholas Wolczynski\thanks{These authors contributed equally to this work.}}
\AFF{Martin Tuchman School of Management, New Jersey Institute of Technology}
\AUTHOR{Yotam Liel\footnotemark[1]}
\AFF{Coller School of Management, Tel Aviv University \\
Cornell Tech, Cornell University}
\AUTHOR{Maytal Saar-Tsechansky}
\AFF{McCombs School of Business, University of Texas at Austin}
\AUTHOR{Lior Zalmanson}
\AFF{Coller School of Management, Tel Aviv University}
} % end of the block
\date{\today}

\begin{document}

\ABSTRACT{
AI advising systems are typically designed to always provide a response. Recent work has focused on developing systems that withhold advice in some instances, such as when advice quality is questionable, to improve the overall quality of guidance when it is given; a design paradigm we refer to as AI abstention. While substantial work has focused on optimizing when models should abstain, little is known about how users respond to systems that abstain. In a preregistered experiment, participants made repeated financial decisions, assisted by one of two advising regimes: an AI that advises selectively or one that advises on every decision. We find that the same AI abstention regime produces differential effects depending on the user: it increases advice acceptance among users with high perceived expertise and tends to decrease it among users with low perceived expertise. This differential effect is accounted for by the cognitive burden users face when repeatedly reconciling AI recommendations with their own judgment; AI abstention appears to relieve this burden among users with high perceived expertise, and in doing so increases their receptivity to advice. Further, AI abstention mitigates the decline in advice acceptance observed among expert users over repeated interactions. Exploratory evidence also suggests it increases appropriate acceptance where advice would improve the decision, increasing the downstream accuracy of perceived experts. The AI abstention regime is thus a first-order design property of AI advisory systems, and one that must be calibrated to the users a system serves.}

\KEYWORDS{AI abstention; human-AI collaboration; advice taking; cognitive burden; personal finance}
%\HISTORY{}

\maketitle

\section{Introduction}
A main challenge for both predictive and generative AI advising systems is that they can produce incorrect responses, including misclassifications, erroneous decisions, and hallucinations. These failures arise in part because AI systems are predominantly optimized to always provide a response even when that response is unlikely to be correct \citep{KalaiEtAl2025_ArXiv_WhyLLMsHallucinate}. Recent work, both from academia and from frontier labs such as OpenAI \citep{KalaiEtAl2025_ArXiv_WhyLLMsHallucinate, Feng2024_ACL_DontHallucinate, fanconi2026cascaded, ZhangEtAl2023_ArXiv_RTuning}, has focused on developing systems that offer advice selectively, withholding advice in some instances to improve the overall quality of guidance when advice is given.
%We formally define AI abstention as an AI's ability to abstain from giving a response; when deployed as an AI abstention policy dictating when the AI abstains from offering advice, this ability confronts users with a user-facing abstention regime — a point we develop below. 
AI designers may want AI models to abstain when the response is unlikely to be appropriate, where an inappropriate response may be one that is highly uncertain, lacks evidence, or may be harmful to the recipient or to others \citep{kirichenko2026abstentionbench, Wen2025_TACL_KnowYourLimits}. AI abstention encompasses a range of implementation strategies, including predictive methods that calculate optimal conditions for when predictive classification models should provide predictions \citep{NotiChen2023_IJCAI_LearningWhenAdvise, Wolczynski2025_WorkingPaper_ValueAIAdvice} and AI abstention strategies where LLMs decline to respond under certain conditions \citep{Wen2025_TACL_KnowYourLimits, Feng2024_ACL_DontHallucinate, ZhangEtAl2023_ArXiv_RTuning, kuhn2023semantic, fanconi2026cascaded}. 

%A growing body of work is focused on developing AI abstention methods for LLMs,  with work from academia and from frontier labs such as OpenAI \citep{KalaiEtAl2025_ArXiv_WhyLLMsHallucinate} proposing that systems should abstain when their confidence is low or the quality of output is insufficient \citep{Feng2024_ACL_DontHallucinate, fanconi2026cascaded, ZhangEtAl2023_ArXiv_RTuning}. 

Research on AI-assisted decision making has theorized many properties of the advisory artifact that shape user reliance, including the accuracy of its recommendations, the explanations it offers, the uncertainty it communicates, and the way it is framed and presented \citep{Jussupow2021_ISR_MedDiagnosisAI, Kim2024_FAccT_UncertaintyExpression, You03042022, Snyder2026}. This literature, however, has taken as given a fundamental design property: that the system provides advice on every decision. Existing theories of advice taking presuppose the presence of advice, treating the user's problem as one of weighting a recommendation that is always there \citep{SniezekVanSwol2001_OBHDP_TrustConfidenceAdvice, Yaniv2004_OBHDP_KnowledgeAdvice}. AI abstention breaks this assumption, and what changes is more than the occasional absence of a response. 

We formally define \textit{AI abstention} as an AI's ability to abstain from giving a response; a system that advises selectively rather than always is defined by an \textit{AI abstention policy} that determines when and for which instances the AI will abstain from offering advice, and a \textit{user-facing AI abstention regime} in which the system's approach to when it advises is communicated to users, who then experience the system sometimes advising and sometimes abstaining. Within such a regime, an individual abstention is not an isolated null event; users interpret it through the communicated rationale that gives it meaning. The user-facing abstention regime is therefore a first-order design property of the advisory artifact, conceptually distinct from the quality of any advice it provides. 
%How users respond to operating under such a regime, and whether their responses differ systematically, remains an open question.

Users' responses to AI abstention regimes may resolve differently across users: some may welcome an advisor designed to provide advice only under certain conditions (e.g., speak only when its input is expected to be most valuable), while others may experience the same policy as withdrawing guidance they rely on. This tension is especially consequential in repeated decisions, in which an AI advisor is embedded in a decision-support workflow and a knowledgeable user must integrate or override its recommendations across a sequence of related tasks. In such settings, by filtering out unreliable or low-value guidance \citep{Wolczynski2025_WorkingPaper_ValueAIAdvice, NotiChen2023_IJCAI_LearningWhenAdvise, Feng2024_ACL_DontHallucinate, fanconi2026cascaded}, AI abstention reduces the volume of recommendations users must process and evaluate, potentially lowering the cognitive burden of incorporating AI assistance into their decisions \citep{10.1145/3757632, You03042022}. At the same time, some users may prefer to receive advice consistently and evaluate it on their own terms \citep{Dietvorst2018_MgmtSci_ModifyAlgorithm, Adam2024_DSS_AIDelegation}. 

Understanding how AI abstention shapes human-AI collaboration is essential for realizing the benefits of these systems in practice. If users respond negatively, becoming less receptive to recommendations when they are offered, the performance gains from higher-quality advice may be partially or fully offset. 
%These risks are especially pronounced in settings where decision-makers cannot verify the objective accuracy of the AI and must form their attitudes based entirely on the subjective interaction experience \citep{Branson2021_BMJQS_DiagnosisLearningCycle, YanivKleinberger2000_OBHDP_Reputation}. 
Further, if the effects of AI abstention vary across users, with AI abstention increasing receptivity for some and reducing it for others, then effective deployment requires understanding which users respond positively and which do not, so that systems can be designed and personalized accordingly. This may be particularly consequential for expert users in high-stakes domains such as medical diagnosis and financial advising, where under-reliance on superior AI recommendations is already well documented\citep{Logg2019_OBHDP_AlgorithmAppreciation, Jussupow2021_ISR_MedDiagnosisAI}, and where improving receptivity without imposing additional cognitive burden could have meaningful practical impact \citep{ancker2017, lebovitz_engage_2022}.

To address these challenges, we ask how AI abstention, implemented as a user-facing AI abstention regime in the context of LLMs, affects advice acceptance. More specifically, we ask whether these effects vary systematically by perceived expertise, whether AI abstention reduces cognitive burden, and whether this reduction is associated with increased receptivity to advice. 

We study these questions with an online experiment in which participants make a series of personal finance decisions with AI assistance, reflecting how AI decision support is often deployed for expert decision-makers \citep{Yang2026, lebovitz_engage_2022}. Participants were randomly assigned to one of two conditions: one in which the AI abstains for a subset of decisions, and a control condition, in which the AI always provides advice. Importantly, participants never learn whether they or the AI was correct, mirroring real-world settings where feedback is absent or substantially delayed. We measure advice acceptance as the primary outcome, estimate the moderating effect of perceived expertise, and and examine whether cognitive burden is a pathway associated with these effects.

%We find a significant crossover interaction between AI abstention and perceived financial expertise: AI abstention increases advice acceptance among users with high perceived expertise and tends to decrease it among users with low perceived expertise. This effect emerges dynamically over the course of the interaction: participants with high perceived expertise in the control condition accept less advice as the engagement progresses, while those who receive advice selectively maintain consistent acceptance. This pattern is consistent with field evidence showing that expert users may reduce reliance on AI systems over time, sometimes leading to abandonment \citep{Jussupow2021_ISR_MedDiagnosisAI, Allen2022_OrgSci_AlgorithmAugmentedWork, lebovitz_engage_2022}. AI abstention appears to mitigate this decline. We further find that AI abstention reduces cognitive burden for users with high perceived expertise, and that this reduction provides evidence for a burden-reduction pathway that accounts for their increased acceptance. These results suggest that by lowering the evaluative cognitive load these users experience from reconciling AI recommendations, AI abstention helps to sustain their engagement rather than eroding it. 

This study makes three contributions to research on human-AI collaboration in information systems. First, we theorize and and empirically examine the behavioral consequences of a user-facing AI abstention regime. Our treatment is the user-facing regime itself -- the experience of a system that abstains from offering advice for some instances combined with a clearly communicated selective-advising policy -- rather than any particular abstention rule, uncertainty estimator, or advising model. By applying AI abstention randomly, we ensure that users' responses reflect the experience of the regime rather than features of a specific policy. Additionally, because our design holds objective advice quality constant across conditions, the effects we document cannot be attributed to better advice. They reflect users' behavioral responses to the regime as an interaction experience, extending theories of advice-taking built on the assumption of ever-present advice to the selective advisors now emerging in practice, and repositioning a choice usually treated as a technical implementation detail as a determinant of user behavior that must be theorized alongside accuracy, explanation, and trust.

Second, we develop and test an expertise-contingent account of how users respond to the AI abstention regime. We find that the effect varies with users' perceptions of their own expertise, producing a crossover in which the same regime can increase advice acceptance among users with higher perceived expertise while tending to decrease it among users with low perceived expertise. This implies that the behavioral consequences of AI abstention cannot be characterized by an average treatment effect, and that theories of reliance on AI must incorporate decision-makers' perceptions of themselves.

%Second, we develop and test an expertise-contingent account of how users respond to AI abstention. Prior work offers two opposing predictions: a selective-advising policy may read as a withdrawal of needed support, eroding receptivity \citep{Wester2024, zheng-etal-2025-easy}, or as relief from the demand of evaluating recommendations, sustaining it. We show that both predictions are correct for different users. Perceived expertise determines which effect prevails, producing a crossover in which the same regime can increase advice acceptance among users with higher perceived expertise and while tending to decrease it among users with low perceived expertise. Notably, both groups operated under identically framed and favorably communicated policies, so the divergence reflects how users respond to the regime, not differences in how it was presented. This implies that the behavioral consequences of AI abstention cannot be characterized by an average treatment effect, and that theories of reliance on AI must incorporate who the user believes themselves to be.

Third, we identify cognitive burden as a mechanism linking the AI abstention regime to sustained engagement, and in doing so offer an explanation for a pattern that prior research has documented but not fully explained: the progressive disengagement of expert users from AI decision support over repeated use \citep{Jussupow2021_ISR_MedDiagnosisAI, lebovitz_engage_2022, ancker2017}. In our study, this decline emerges even though the advice is always objectively appropriate, thus the erosion is driven not by advice failures but by the accumulating cost of evaluating advice. An AI abstention regime interrupts this dynamic, preserving rather than raising expert receptivity by relieving the evaluative demand that continuous advising imposes. Cognitive burden thus emerges as a mechanism through which the under-reliance problem in expert-facing deployments can be addressed.

\section{Related Work \& Theory Development}
\label{sec:literature}
In this section we review relevant literature and introduce the theory that motivates our hypotheses. Figure~\ref{fig:theory} demonstrates the theoretical model we develop, which we later test empirically. 

\subsection{AI abstention}

AI advising systems can improve overall guidance quality by abstaining from providing advice in instances where providing it would be unhelpful or harmful \citep{NotiChen2023_IJCAI_LearningWhenAdvise, Wolczynski2025_WorkingPaper_ValueAIAdvice, Feng2024_ACL_DontHallucinate}. We refer to this paradigm as \emph{AI abstention}, where models are trained or prompted to decline to respond under certain conditions, such as not to produce unreliable or harmful outputs \citep{Wen2025_TACL_KnowYourLimits, Feng2024_ACL_DontHallucinate, fanconi2026cascaded}. A substantial body of work has emerged on how to develop and optimize such systems. On the classification side, researchers have developed learning-based policies that determine when algorithmic classification predictions would benefit human decision-makers, deferring to human judgment when the algorithm is unlikely to improve on it \citep{NotiChen2023_IJCAI_LearningWhenAdvise}, and demonstrated that personalizing abstention to a decision-maker's individual strengths produces advice more likely to improve on their initial judgment \citep{Wolczynski2025_WorkingPaper_ValueAIAdvice}. In the LLM context, substantial work has focused on training models to recognize and express uncertainty \citep{Kadavath2022_ArXiv_LanguageModelsKnow, LinEtAl2022_TACL_TeachingUncertainty, kuhn2023semantic}, to abstain when uncertain \citep{Feng2024_ACL_DontHallucinate, ZhangEtAl2023_ArXiv_RTuning}, and to balance coverage against the risk of providing incorrect guidance \citep{Tomani2024_ArXiv_UncertaintyAbstention}. The underlying premise that AI abstention can improve system performance is now well established, and the field has moved toward optimizing the mechanisms of abstention rather than questioning its value \citep{YinEtAl2023_ACL_DoLLMsKnow, Feng2024_ACL_DontHallucinate, fanconi2026cascaded}. This literature concerns the technical mechanism of abstention; in deployment, that mechanism reaches users as a user-facing AI abstention regime which includes both the communication of the technical mechanism guiding the advice provision policy and the experience of that policy in practice, which is the object of the present study.

As argued in the introduction, this technical progress leaves the behavioral side of AI abstention undertheorized: optimizing when a model abstains is necessary but not sufficient, because the benefits of higher-quality advice can only be realized if users remain receptive to the advice they do receive. A small literature has begun to examine user responses to AI abstention, though no existing work addresses our setting directly. Studies of safety-driven refusals find that refusal messages are generally not favorably perceived, leading to frustration \citep{Wester2024, zheng-etal-2025-easy}. Work on uncertainty expression finds that first-person expressions of uncertainty can reduce over-reliance in information retrieval settings \citep{Kim2024_FAccT_UncertaintyExpression}. Crucially, these works focus on safety guardrail responses or factual information retrieval, not uncertain decision-making domains without immediate feedback. Further, none of it examines downstream behavioral outcomes such as advice acceptance nor considers whether user responses differ across individuals.

%This gap is consequential. If users respond negatively to AI abstention in uncertain decision-making settings, the performance gains from higher-quality advice may be offset by reduced receptivity. And if these responses vary across users, a single deployment strategy risks helping some users while harming others. We address this gap directly, examining how users respond to AI abstention in an uncertain decision-making task without immediate feedback, and whether these responses differ by perceived expertise.

\subsection{Expertise-Contingent Reliance on AI Advice}

When people collaborate with advisors, patterns of reliance vary systematically with expertise. Experts tend to exhibit under-reliance, discounting even superior recommendations in favor of their own judgment, while novices tend to over-rely, following even low-quality advice without sufficient scrutiny \citep{PassiVorvoreanu2022_MSRTR_Overreliance, Nourani2020_HCOMP_DomainExpertiseTrust, Liel02102025}. In AI contexts specifically, domain expertise reduces trust in decision support systems and decreases the likelihood of following recommendations \citep{Bayer2022_JDecisSyst_XAIExpertise, Jussupow2021_ISR_MedDiagnosisAI, Nourani2020_HCOMP_DomainExpertiseTrust}, with experienced radiologists more likely to ignore correct AI advice, especially when it disconfirms their initial diagnosis, while novices are more susceptible to being misled by incorrect recommendations \citep{Jussupow2021_ISR_MedDiagnosisAI}. The subjective confidence a person has in their own competence (their perception of their expertise) is the primary driver of this pattern: those who feel more capable are less willing to defer to outside guidance \citep{SeeEtAl2011_OBHDP_PowerConfidence, GinoMoore2007_JBDM_TaskDifficulty, Yaniv2004_OBHDP_KnowledgeAdvice}. Because this subjective sense of competence is our focal construct, we operationalize expertise as perceived expertise throughout. For brevity, we refer to users with high and low perceived expertise as experts and non-experts; this is terminological shorthand for the perceived-expertise construct and carries no implication that participants hold credentialed or professional expertise. 

These divergent patterns have meaningful and well-documented consequences. Expert under-reliance means missing the benefits of superior recommendations: professional forecasters with domain expertise show significantly reduced algorithm appreciation compared to laypeople, paradoxically hurting their accuracy \citep{Logg2019_OBHDP_AlgorithmAppreciation, Onkal2009_JBDM_ForecastAdjustments}, and experienced radiologists who dismiss AI advice are more likely to arrive at incorrect diagnoses \citep{Jussupow2021_ISR_MedDiagnosisAI}. Over-reliance, in contrast, means accepting inferior recommendations without the judgment to recognize when the AI is wrong \citep{PassiVorvoreanu2022_MSRTR_Overreliance}. Both patterns undermine the value that AI advising systems are designed to deliver \citep{Vaccaro2024_NatHumBehav}. Compounding this, in many of the domains where AI advising systems are deployed, feedback on decision quality is substantially delayed or altogether absent. Physicians often do not receive outcome feedback on diagnostic decisions \citep{Branson2021_BMJQS_DiagnosisLearningCycle, Omron2018_AEMEducTrain_CalibrationGap}, and financial advisors may not observe the outcomes of investment decisions for months or years \citep{Chalmers2020_JFE_ConflictedAdvice}. Without timely feedback, users cannot verify whether provided advice was correct or whether advice that was abstained would have been helpful. This means that attitudes toward AI abstention are shaped entirely by the subjective experience of the interaction, making it especially important to understand how that experience differs across users with varying levels of perceived expertise.

%Understanding whether AI abstention affects decision-makers that perceive themselves to be experts and non-experts differently is an open and important question. 
If AI abstention increases advice acceptance among users with high perceived expertise, and given that professionals tend to also perceive themselves to be experts, it could address one of the most persistent challenges in human-AI collaboration: getting professionals to rely more appropriately on AI recommendations in domains where under-reliance is well documented. If AI abstention reduces acceptance among users with low perceived expertise, it risks reducing receptivity to guidance among users who actively seek and depend on consistent AI recommendations. We address these open questions directly, treating perceived expertise as the moderator of the AI abstention effect and examining the cognitive mechanism through which this moderation operates.

\subsection{Cognitive Burden of Reconciling AI Advice}

Processing advice places real cognitive demands on users \citep{Yaniv2004_OBHDP_KnowledgeAdvice, SollLarrick2009_JEPLMC_RevisingJudgment}. Evaluating a recommendation, weighing it against one's own judgment, and deciding whether to act on it requires active mental effort \citep{Yaniv2004_OBHDP_KnowledgeAdvice, Todd1992, Todd1994}. An advisor’s presence also adds a decision that would not otherwise exist -- whether to consult, follow, or disregard its recommendation \citep{Snyder2026}. When these demands accumulate across repeated interactions, they can erode users' willingness to engage with advice altogether \citep{10.1145/3757632, lebovitz_engage_2022}. Drawing on cognitive load theory \citep{Sweller1988}, we refer to this demand as cognitive burden. We use cognitive burden rather than cognitive load because our focus is the recurring cost of processing and reconciling AI advice, which accumulates across repeated interactions. Field evidence from professional settings illustrates this clearly: in a study of diagnostic radiologists using AI tools, professionals in two of three departments found the cognitive demands of engaging with AI recommendations so burdensome that they disengaged from AI inputs entirely, failing to incorporate them into their final judgments \citep{lebovitz_engage_2022}. Research on clinical decision support systems similarly finds that high volumes of recommendations requiring cognitive evaluation produce alert fatigue that substantially reduces acceptance rates \citep{ancker2017}. More broadly, elevated cognitive burden has been shown to reduce compliance with recommendations across a range of settings \citep{Wogalter1999, Wogalter1998}, and research on algorithmic advice specifically finds that when processing AI recommendations imposes greater cognitive burden, advice taking declines \citep{You03042022}.

Cognitive burden is an important outcome to consider in its own right. Expert users in professional settings already operate under substantial cognitive demand, and AI advising systems that compound this burden risk degrading the quality of human-AI collaboration regardless of the quality of the advice being offered \citep{lebovitz_engage_2022, ancker2017}. Yet despite its practical importance, cognitive burden has received relatively little attention as an outcome variable in research on AI advising system design. Prior work has largely focused on perceptions, trust, and performance as the key outcomes of human-AI collaboration, leaving open the question of whether system design choices that improve these outcomes do so at the cost of increased cognitive demand on users. We examine cognitive burden as both a consequential outcome and a mediating mechanism, asking whether AI abstention relieves the cognitive demands placed on users and whether this relief differs across users with varying levels of perceived expertise.

We theorize that cognitive burden operates as a moderated mediator, with AI abstention reducing burden differently depending on users' perceived expertise, and this differential reduction in turn accounting for the effect on advice acceptance. For users with high perceived expertise, the cognitive burden of processing AI advice stems in part from having to repeatedly evaluate whether to trust and defer to recommendations that conflict with their confident self-assessment \citep{Yaniv2004_OBHDP_KnowledgeAdvice, SeeEtAl2011_OBHDP_PowerConfidence}. AI abstention reduces the volume of such evaluations, potentially relieving a meaningful portion of this burden. For users with low perceived expertise, the calculus is different. These users perceive the underlying tasks themselves as cognitively demanding and rely on AI guidance to reduce the burden of working through unfamiliar problems. Therefore, when advice is absent, low-expertise users are not expected to experience relief in cognitive burden, as they are left facing the decision itself as well as the interpretation of the system's decision not to advise. As a result, we theorize that AI abstention will reduce cognitive burden for users with high perceived expertise but not necessarily for users with low perceived expertise, and this differential reduction is a pathway through which AI abstention increases advice acceptance for the former group but not the latter.

\subsection{Hypotheses}
\begin{figure}[t]
\FIGURE
{\includegraphics[width=0.8\textwidth]{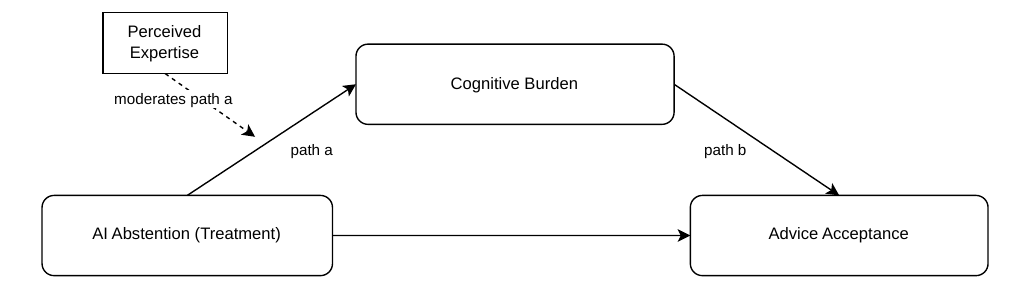}}
{Theoretical Model of AI Abstention Effects on Advice Acceptance.\label{fig:theory}}
{AI abstention affects advice acceptance through cognitive burden (paths a and b). Perceived financial expertise moderates the effect of AI abstention on cognitive burden (path a).}
\end{figure}

Figure~\ref{fig:theory} presents our theoretical model. AI abstention affects advice acceptance both directly and through its effect on cognitive burden, with perceived financial expertise moderating the path from AI abstention to cognitive burden.
%\begin{figure}[t]
%\centering
%\includegraphics[width=0.97\textwidth]{Figures/selective_theory_diagram.pdf}
%\caption{Theoretical Model of AI Abstention Effects on Advice Acceptance. \textit{Note.} AI abstention affects advice acceptance through cognitive burden (paths a and b). Perceived financial expertise moderates the effect of AI abstention on cognitive burden (path a).}
%\label{fig:theory}
%\end{figure}

Building on this theoretical model, we derive three hypotheses. Our first hypothesis concerns the moderated effect of AI abstention on advice acceptance. As reviewed above, reliance on advice differs between users who perceive themselves as expert and those who do not: users who perceive themselves as expert tend to under-rely on recommendations, discounting advice in favor of their own judgment, whereas users with lower perceived expertise depend more heavily on guidance. For high-perceived-expertise users, AI abstention aligns with their preference to engage with advice selectively; for low-perceived-expertise users, the same selectivity withdraws guidance they rely on. We therefore expect the effect of AI abstention on advice acceptance to depend on perceived expertise, with AI abstention having a more positive effect on advice acceptance for users with high perceived expertise than for users with low perceived expertise.

\begin{itemize}
    \item \textbf{H1:} The effect of AI abstention on advice acceptance is moderated by perceived financial expertise, such that the effect is more positive for users with high perceived expertise than for users with low perceived expertise.
\end{itemize}

The moderation may appear as a crossover, with abstention leading to a negative effect for non-expert users. Our prediction, however, concerns the difference between groups rather than the sign of the effect within either one.

Our second hypothesis concerns the effect of AI abstention on cognitive burden. For users with high perceived expertise, the burden of processing AI advice stems in part from having to repeatedly evaluate whether to trust and defer to recommendations that may conflict with their confident self-assessment. AI abstention reduces the volume of such evaluations, potentially relieving a meaningful portion of this burden. For users with low perceived expertise, however, reducing the volume of advice is not expected to relieve burden: these users perceive the underlying tasks as cognitively demanding and rely on AI guidance to ease them, so AI abstention leaves them facing both the decision itself and the interpretation of the system's decision not to advise.

\begin{itemize}
\item \textbf{H2:} The effect of AI abstention on cognitive burden is moderated by perceived financial expertise, such that AI abstention reduces cognitive burden for users with high perceived expertise more than for users with low perceived expertise.
\end{itemize}

Our third hypothesis concerns the mediating role of cognitive burden. If AI abstention reduces cognitive burden for users with high perceived expertise, and if lower cognitive burden in turn increases advice acceptance, then cognitive burden is a pathway through which AI abstention differentially affects acceptance across expertise groups.

\begin{itemize}

\item \textbf{H3:} The moderated effect of AI abstention on advice acceptance operates in part through cognitive burden, such that the indirect effect is more positive for users with high perceived expertise than for those with low perceived expertise.
\end{itemize}

\section{Experiment}
\label{sec:experiment}

We conducted a pre-registered online experiment examining how AI abstention affects advice acceptance across a series of personal financial decision-making tasks. 

\subsection{Study Materials}

\subsubsection{Task Design and Development}

We developed 10 financial decision scenarios covering investment allocation, credit card selection, retirement planning, stock valuation, debt management, and estate planning. Financial decisions provide an accessible analog for the kinds of high-stakes uncertain decision-making settings -- medical diagnosis, financial advising, professional forecasting -- where AI abstention is most consequential, while allowing us to recruit a broad participant pool with varying levels of self-assessed expertise \citep{Kleinberg2018_QJE_HumanMachinePredictions}. Each scenario presents a concrete financial situation with four to five choice options following a dual-optimal structure: two options are designated as superior based on established financial principles, while the remaining options are clearly inferior. This structure reflects real-world decision ambiguity where multiple good solutions exist and decision makers cannot definitively verify which specific option is ``best'' (see Online Appendix~\ref{appendix:scenarios} for complete scenario details).

The dual-optimal design enables experimental control over AI agreement and disagreement independent of participant performance. When a participant selects a superior choice, the AI can either confirm that choice (agreement) or recommend the alternative superior choice (recommend alternative). Scenarios were validated through manual review and validation with multiple large language models to ensure reliable identification of the two superior options.

\subsubsection{AI Advice}

For each possible participant choice, we generated structured AI explanations using Claude Sonnet 4.5. Explanations follow the style of real-world AI advisory tools, providing the relevant financial principles and analysis before concluding with a recommendation. AI explanations appear to be progressively typed, simulating how language models generate text, mimicking realistic AI advisory tools and supporting participant engagement with the advice. See Online Appendix~\ref{appendix:ai} for full explanation generation protocol and example responses.

\subsubsection{Instruction Pre-Testing}

Given that AI abstention is a relatively novel mechanism, we conducted an iterative pre-test to develop instructions that clearly conveyed this advising approach. We recruited a total of 360 participants who read different versions of the instructions and then completed comprehension questions and ratings of clarity and credibility. We refined the instructions across iterations until they reached high levels of comprehension, clarity, and credibility.

\subsection{Experiment Design and Procedure}

\subsubsection{Overview}

\begin{figure}[t!]
\FIGURE
{\includegraphics[width=0.8\linewidth]{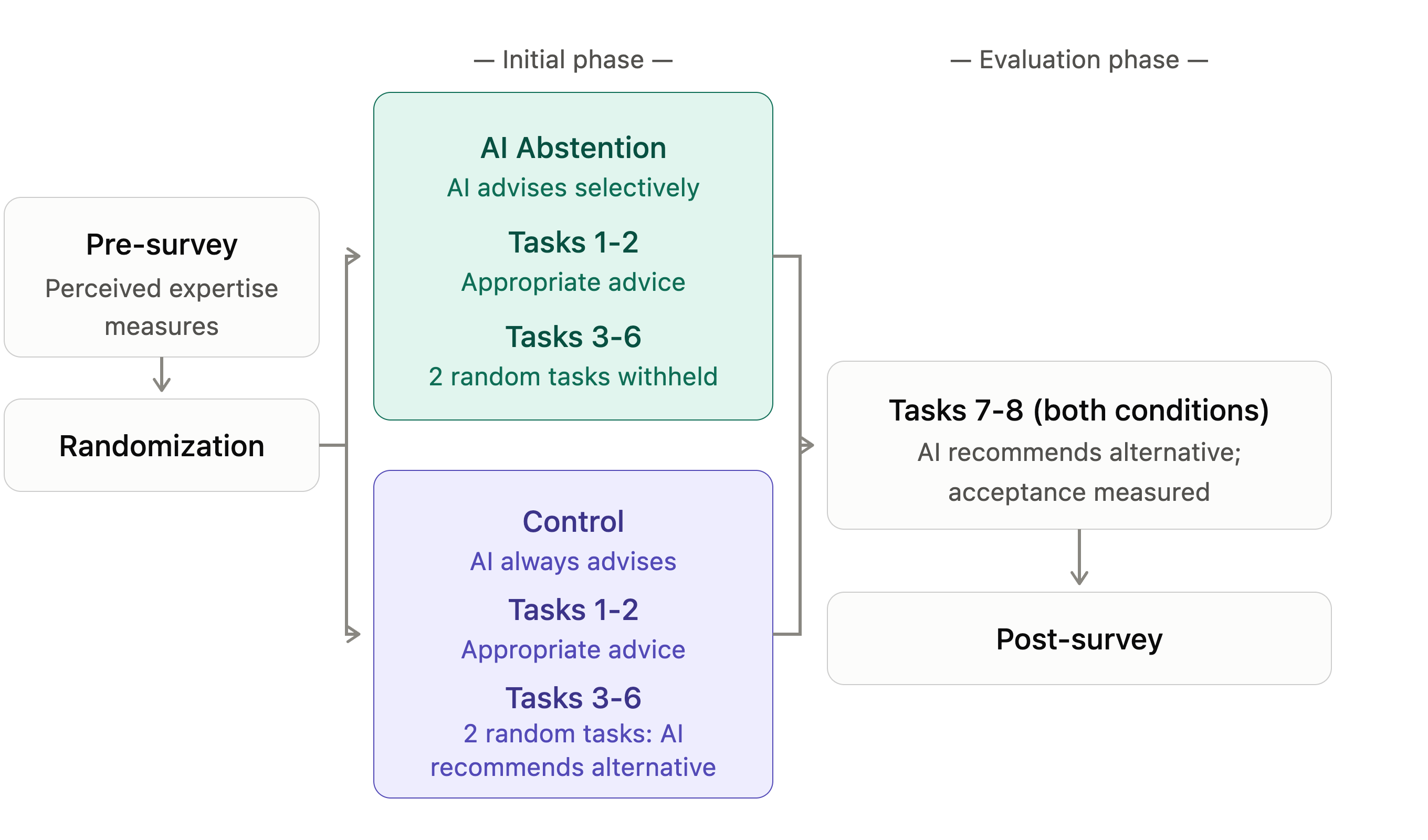}}
{Experiment design.\label{fig:design}}
{Note. Participants completed a pre-survey measuring perceived financial expertise, then were randomly assigned to condition. AI abstention withheld advice on two randomly selected tasks during the treatment portion of the initial phase (tasks 3–6); the control condition advised on all tasks, with two randomly selected tasks receiving a recommendation for the alternative option. Both conditions received identical advice in the evaluation phase (tasks 7–8), where advice acceptance was measured, followed by a post-survey.}

\end{figure}

Participants were randomly assigned to one of two conditions: an AI abstention treatment condition, and a control condition where advice was provided on every task. The randomization was stratified by perceived expertise to ensure balance across groups. Participants completed eight decisions with AI assistance: we manipulated AI advice provision during an initial six-task phase and measured downstream effects on advice acceptance in a subsequent two-task evaluation phase. Participants received no feedback on decision correctness throughout, mirroring real-world professional settings where feedback is absent or substantially delayed and where attitudes toward AI systems must be formed based entirely on the subjective interaction experience. A diagram of the experiment design is shown in Figure \ref{fig:design}.

Participants began by completing a brief pre-survey that included a financial expertise scale, adapted from the Consumer Financial Protection Bureau's Financial Skill Scale \citep{CFPB_FinancialSkillScale}, along with a single-item perceived financial expertise question on a 5-point Likert scale used to stratify treatment assignment and ensure similar distributions of perceived expertise across conditions. Following treatment assignment, participants read condition-specific task instructions and completed an instruction comprehension check to ensure they understood whether the AI would abstain for some decisions or always provide advice. They then completed eight financial decision tasks: six in an initial phase where the treatment was administered, followed by two in an evaluation phase where we measured advice acceptance. Six scenarios were designated for the initial phase and presented to all participants in random order. Four scenarios were designated as potential evaluation tasks, and two were randomly selected for each participant and presented in random order. Tasks were presented continuously without explicit phase separation.

Following the eight decision tasks, participants completed a post-experiment questionnaire with an attention check item embedded in the survey. The experiment lasted approximately 15 minutes.

\subsubsection{Task Structure}

\begin{figure}[t!]
\FIGURE
{\includegraphics[width=0.9\linewidth]{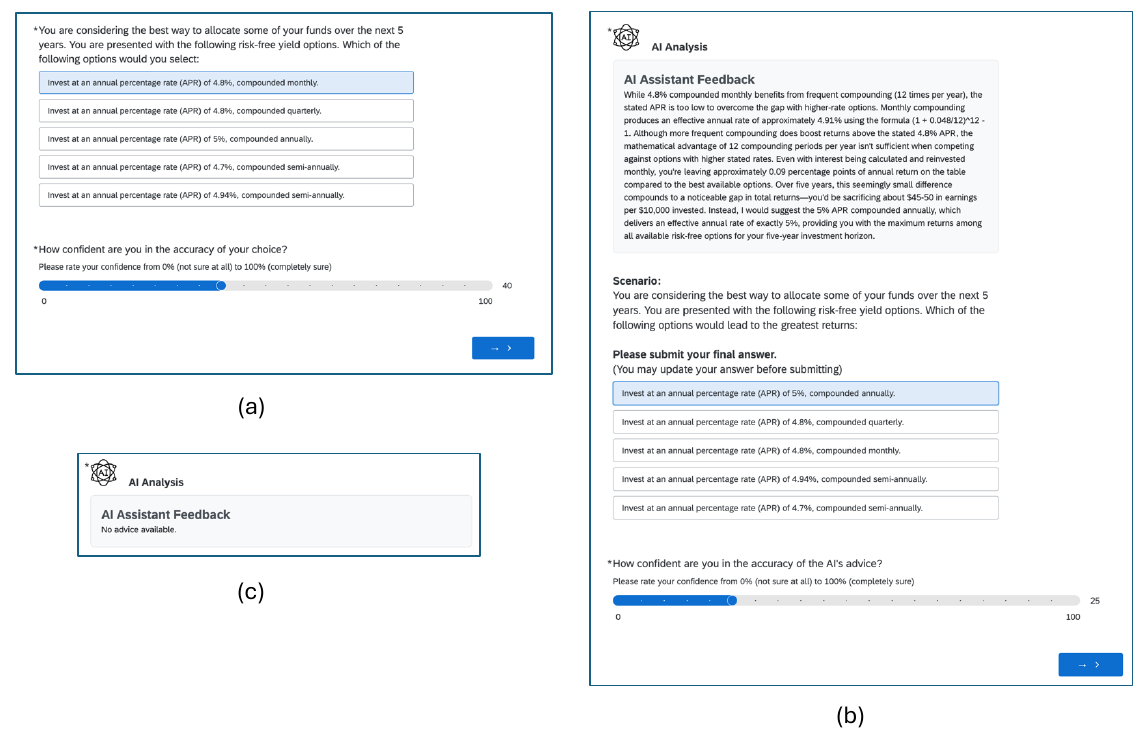}}
{Example decision task as presented to participants.\label{fig:task_example}}
{(a) Shows interface before receiving AI advice, (b) shows interface after receiving AI advice, and (c) shows just the AI response when no advice is provided.}
\end{figure}

Each task followed a judge-advisor system (JAS) paradigm: participants provided an initial judgment and rated their confidence in their choice before being exposed to advice \citep{SniezekVanSwol2001_OBHDP_TrustConfidenceAdvice}. All participants then saw an AI message that appeared progressively on screen. This message was either substantive advice or a brief abstention message reading ``No advice available.'' Participants then saw the decision scenario again with their initial choice pre-filled and could confirm or change their selection. When substantive AI advice was provided, participants additionally rated their confidence in the AI's recommendation. An example question, as displayed before and after receiving AI advice, along with an example of the message shown to participant's when AI advice is not offered, is shown in Figure~\ref{fig:task_example}.

\subsubsection{Defining AI Abstention}

Our treatment is intended to capture the effect of exposure to a user-facing AI abstention regime. As argued in the introduction, users of systems that may abstain do not encounter the abstention in isolation: they are typically given some explanation of the system's advising policy and then experience the system sometimes providing advice and sometimes abstaining. We therefore define the treatment as the combined regime of abstention instructions and subsequent exposure to abstention of advice. We do not treat abstaining alone as a separable object of study because users necessarily interpret non-advice through some explanation of why advice is absent. A system that advises selectively rather than always is therefore defined by both the act of abstaining and the policy rationale that gives that abstaining meaning. Participants encountered this regime under the label "selective advising" (see Online Appendix \ref{appendix:instructions} for the full instructions); throughout the remainder of the paper, we refer to this experimentally implemented user-facing regime as AI abstention. The estimand is the behavioral effect of this regime, rather than the effect of any particular technical abstention rule, uncertainty estimator, or model-selection policy.

\subsubsection{Instructions}

Participants in both conditions were informed they would have access to an AI financial advising model to assist in their decisions. Additionally, the control condition was informed the AI would use a comprehensive advising approach and provide a recommendation on every decision, while the treatment condition was informed the AI would use an AI abstention approach and provide recommendations when its input is judged to be beneficial. Both descriptions used parallel framing, presenting the respective approach as a deliberate design choice intended to support decision quality and listing analogous benefits. Full instructions for both conditions are provided in Online Appendix~\ref{appendix:instructions}. Following the instructions, all participants completed a comprehension check confirming their understanding of whether the AI would abstain on some decisions or provide advice for every decision.

\subsubsection{Treatment (AI Abstention)}
The AI provides advice on six of the eight tasks. During the first two tasks, all participants receive appropriate advice: the AI confirms optimal choices and provides corrective feedback recommending one of the superior options when the participant selects a suboptimal option. During tasks 3 to 6, two tasks are randomly selected for abstention, where the AI displays a brief message reading ``No advice available.'' regardless of the participant's choice. The remaining tasks in this phase provide appropriate advice as in the control condition.

We randomize which tasks the system abstains on to identify the behavioral effect of experiencing a user-facing selective-advising regime, independent of any particular technical abstention rule. Real AI abstention systems may withhold advice for many reasons, including model uncertainty, predicted value of advice, task difficulty, user characteristics, or some combination of these factors. If abstaining were tied to one such rule in our experiment, users’ responses could reflect features of that rule rather than the general experience of interacting with a system that sometimes provides advice and sometimes withholds it. Randomizing abstaining therefore allows us to study the user-facing experience of selectivity without confounding it with a specific algorithmic policy.

\subsubsection{Control (Always Advise)} The AI provides advice on all eight tasks, reflecting the status quo of systems that always provide recommendations. The first two tasks follow the same structure as the treatment. During tasks 3 to 6, the design parallels the treatment with one substitution: rather than abstaining, two randomly selected tasks instead receive a recommendation to choose the alternative superior option, regardless of the participant's choice. This advice remains accurate because both options are appropriate, but is sometimes less directly useful such as if the user's initial choice was already appropriate, reflecting the kind of recommendation an AI abstention mechanism would naturally suppress.

\subsubsection{Evaluation Phase} The evaluation phase (tasks 7 and 8) was identical in both conditions: the AI always recommends one of the superior options, regardless of the participant's initial choice, recommending the alternative superior option when the participant selects the other superior choice and one of the two superior options when the participant selects a suboptimal choice. This design ensures advice acceptance is measured under comparable conditions for all participants regardless of initial choice quality. In our analyses, we control whether participants had initially chosen one of the superior options and assess whether there were differences in outcomes depending on the user's initial choice.

Our design also holds objective advice quality constant across conditions. The AI always provides objectively correct advice in either condition. In the treatment condition, abstaining sometimes suppresses advice that would have been accurate; in the control condition, the AI provides that advice. Thus, any treatment effect cannot be attributed to AI abstention providing more accurate advice than always advising. Instead, the estimated effect reflects how users respond to the experienced abstention regime, holding objective advice correctness constant. In real deployments, an AI abstention regime may also improve average advice quality by abstaining from providing low-confidence or low-value recommendations. Our design deliberately abstracts away from that performance channel so that we can isolate the behavioral response to the user-facing AI abstention regime as an interaction experience.

\subsection{Measures}

\subsubsection{Per-Task Measures}

For each decision task, participants first select one of the presented options and then rate their confidence in their choice on a scale from 0 to 100. After submitting their choice, they see the AI's response or abstention message and have the opportunity to update their selection. When substantive AI advice was provided, participants additionally rate their confidence in the AI's recommendation on a scale from 0 to 100. Our primary outcome variable is advice acceptance, measured during the evaluation phase when both conditions receive AI recommendations for one of the superior choices. Advice acceptance equals 1 if the participant updates their initial choice to the AI-recommended option; 0 otherwise.

\subsubsection{Post-Survey Measures}

Following all decision tasks, participants complete a post-survey measuring perceived cognitive burden using three items on a 5-point scale, adapted from a validated cognitive load measure \citep{KlepschSchmitzSeufert2017_FrontPsych_CognitiveLoadInstruments}, along with a suite of additional measures collected for exploratory purposes, see Appendix~\ref{appendix:scales} for complete scale items and Cronbach Alpha values. We follow standard experimental design practice and collect these measurements in a post-survey to ensure the questionnaire does not prime participants during the task and consequently impact the primary dependent variable, advice acceptance \citep{Yin2021, Adam2025-rp, Burke2025}. We thus use the term cognitive burden to capture participants’ subjective perception of the mental demand they experienced during the tasks.

\section{Results}
\subsection{Pilot Study and Power Analysis}

Prior to the main experiment, we conducted a pilot study (N = 113) using the experimental design described in Section 3. The pilot was not powered for hypothesis testing; its purpose was to assess the feasibility of the design and to inform a power analysis for the main experiment. The directional pattern across all primary outcomes was consistent with our hypotheses, and the pilot estimates were used to determine the sample size for the pre-registered main experiment. The results reported below therefore constitute a pre-registered test of patterns first observed in an independent sample.

\subsection{Participants}

We recruited 600 participants via Prolific. Participants were excluded for failing automated bot checks, failing the attention check embedded in the post-survey, or failing the comprehension check confirming their understanding of the condition instructions. These exclusions were pre-registered before data was collected and we provide full post-exclusion randomization diagnostics in Appendix~\ref{app:randomization}. This yielded a final analytic sample of 455 participants ($N_{\text{control}} = 213$; $N_{\text{treatment}} = 242$), contributing 910 evaluation phase observations ($Q_{\text{control}} = 426$; $Q_{\text{treatment}} = 484$) and 3,640 total observations across all eight decision tasks. 

We report full descriptive statistics broken down by expertise group, experimental phase, and condition in Appendix \ref{appendix:recruitment}. Participants are classified as perceived experts if their perceived financial expertise score, computed as the mean of their three responses to the adapted Financial Skill Scale items, falls in the top 30\% of the sample ($\text{mean perceived financial expertise} \geq 3.667$, $N = 115$); the remaining 340 participants are classified as non-experts. For readability, we refer to these two subgroups as “Experts” and “Non-Experts” in tables, figures, and subsequent analyses. 
%The table distinguishes three phases of the experiment: the pre-treatment phase (Questions 1--2), in which all participants received appropriate advice regardless of condition; the during phase (Questions 3--6), in which the treatment manipulation was administered; and the evaluation phase (Questions 7--8), in which both conditions received recommendations for one of the superior choices and advice acceptance was measured. 

\subsection{Moderation of Advice Acceptance by Perceived Expertise}
\begin{figure}[t!]
\FIGURE
{\includegraphics[width=0.6\textwidth]{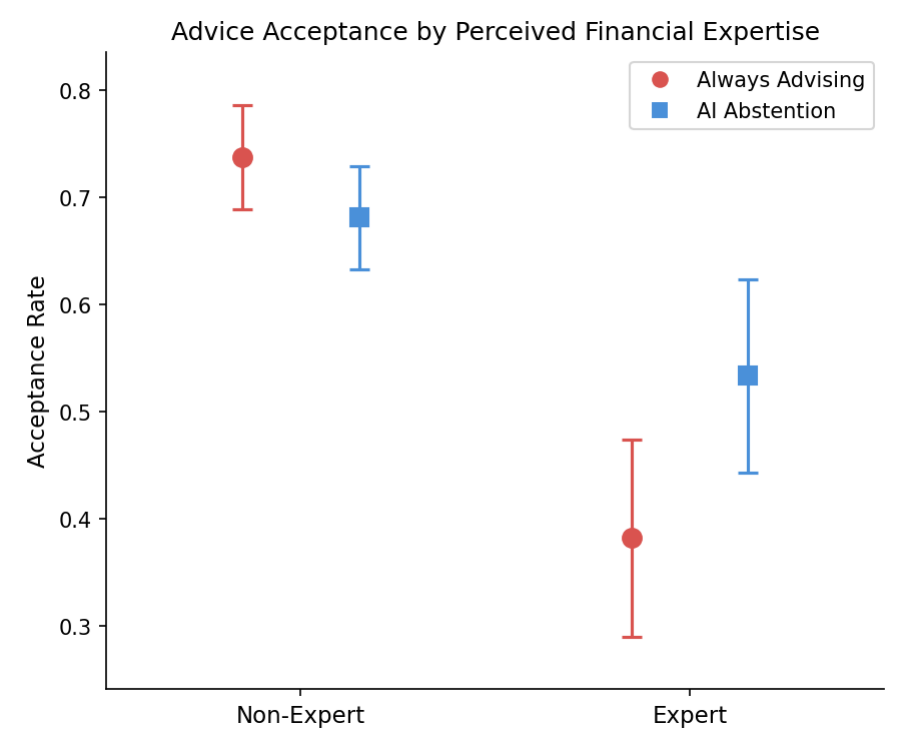}}
{Advice acceptance rates by condition during the evaluation phase.\label{fig:main_effect}}
{Error bars represent 95\% confidence intervals.}
\end{figure}

Figure~\ref{fig:main_effect} displays advice acceptance rates by condition and perceived expertise group during the evaluation phase. The pattern is consistent with H1: AI abstention increases advice acceptance for expert users (38.2\% in control vs. 53.3\% in treatment, +15.1 percentage points) and decreases it for non-expert users (73.4\% in control vs. 68.1\% in treatment, $-$5.3 percentage points), yielding a clear crossover interaction.

%\begin{figure}[t!]
%    \centering
%    \includegraphics[width=0.8\textwidth]{Figures/fig1_main_effect.pdf}
%    \caption{Advice acceptance rates by condition during the evaluation phase. Error bars represent 95\% confidence intervals.}
%    \label{fig:main_effect}
%\end{figure}

To formally test H1, we estimate a logistic regression model of the form:
\begin{equation}
    \Pr(\text{Accept}_{ij} = 1) = F\left(\beta_0 + \beta_1 \text{Treatment}_j + \beta_2 \text{Expertise}_j + \beta_3 \text{Treatment}_j \times \text{Expertise}_j + \mathbf{X}_{ij}'\boldsymbol{\gamma}\right)
    \label{eq:h1}
\end{equation}

\noindent where $F(\cdot)$ denotes the logistic function; $\text{Accept}_{ij}$ is a binary indicator equal to one if participant $j$ updated their initial choice to the AI-recommended option on evaluation task $i$; $\text{Treatment}_j$ is a binary indicator for assignment to the AI abstention condition; $\text{Expertise}_j$ denotes perceived financial expertise, operationalized either as a continuous measure (mean of Financial Skill Scale items) or as a binary indicator for the top 30\% of the expertise distribution; $\text{Treatment}_j \times \text{Expertise}_j$ is the interaction between treatment assignment and perceived expertise; and $\mathbf{X}_{ij}$ is a vector of controls including initial confidence, log transformed initial response time before seeing AI advice, log transformed final response time after seeing the AI advice, question fixed effects, question view order, and a binary indicator of whether the user initially chose one of the two superior options. Standard errors are clustered at the participant level to account for the non-independence of observations within participants. We estimate four models: continuous and binary operationalizations of expertise, each with and without controls.

\begin{table}[t!]
\TABLE
{Logistic Regression: Moderation of AI Abstention Effect by Perceived Expertise\label{tab:h1}}
{%
\begin{tabular}{lcccc}
\toprule
 & Model 1 & Model 2 & Model 3 & Model 4 \\
\midrule
\textbf{Treatment (AI abstention)}                      & $-$0.264  & $-$0.272  & $-$1.606$^{**}$   & $-$1.506$^{*}$ \\
                                                             & (0.228)   & (0.214)   & (0.800)           & (0.794) \\
\addlinespace
Perceived Expert (binary)                                    & $-$1.113$^{***}$ & $-$1.514$^{***}$ &          & \\
                                                             & (0.318)          & (0.296)          &          & \\
\addlinespace
Perceived Expertise (cont.)                                  &           &           & $-$0.732$^{***}$ & $-$0.973$^{***}$ \\
                                                             &           &           & (0.185)          & (0.180) \\
\addlinespace
\textbf{Treatment $\times$ Perceived Expert (binary)}        & \textbf{1.047$^{**}$} & \textbf{0.888$^{**}$} & & \\
                                                             & \textbf{(0.436)}      & \textbf{(0.403)}      & & \\
\addlinespace
\textbf{Treatment $\times$ Perceived Expertise (cont.)}      &           &           & \textbf{0.500$^{**}$} & \textbf{0.450$^{**}$} \\
                                                             &           &           & \textbf{(0.233)}     & \textbf{(0.229)} \\
\addlinespace
Controls                                                     & Yes       & No        & Yes       & No \\
Question fixed effects                                       & Yes       & No        & Yes       & No \\
\addlinespace
\midrule
Observations                                                 & 910       & 910       & 910       & 910 \\
Pseudo $R^2$                                                 & 0.168     & 0.044     & 0.176     & 0.069 \\
\bottomrule
\end{tabular}%
}%
{\NOTE{Notes.}Log-odds from logistic regression, participant-clustered SEs in parentheses. Treatment coefficient reflects the effect of AI abstention for non-experts (reference group). Controls: initial correctness, confidence, log initial and final response time, and view order. $^{*}p < .10$, $^{**}p < .05$, $^{***}p < .01$.}
\end{table}

Table~\ref{tab:h1} reports the results. Our primary specification uses the binary perceived expertise indicator (Models 1 and 2), as we are most interested in the subset of users who perceive themselves as genuinely expert, rather than simply above the median on perceived expertise. The Treatment $\times$ Perceived Expert interaction is positive and significant in both the model without controls (Model 2: $\beta = 0.888$, $p = 0.027$) and the model with controls (Model 1: $\beta = 1.047$, $p = 0.016$), indicating that AI abstention increases advice acceptance for perceived experts relative to non-experts. The treatment coefficient, which captures the effect of AI abstention for non-expert users specifically, is negative in both models, consistent with a crossover pattern. Models 3 and 4 replace the binary indicator with the continuous perceived expertise score as a moderator. The Treatment $\times$ Perceived Expertise interaction remains positive and significant in both Models 3 ($\beta = 0.500$, $p = 0.032$) and 4 ($\beta = 0.450$, $p = 0.049$), demonstrating that the effect is not driven by the choice of binary cutoff. Re-estimating Model 1 at distinct cutpoints on the expertise scale further shows that the interaction is positive and significant at all thresholds at or above the one we adopt, and grows in magnitude as the threshold rises (Appendix~\ref{app:cutoffs}).

\begin{figure}[t!]
\FIGURE
{\includegraphics[width=\textwidth]{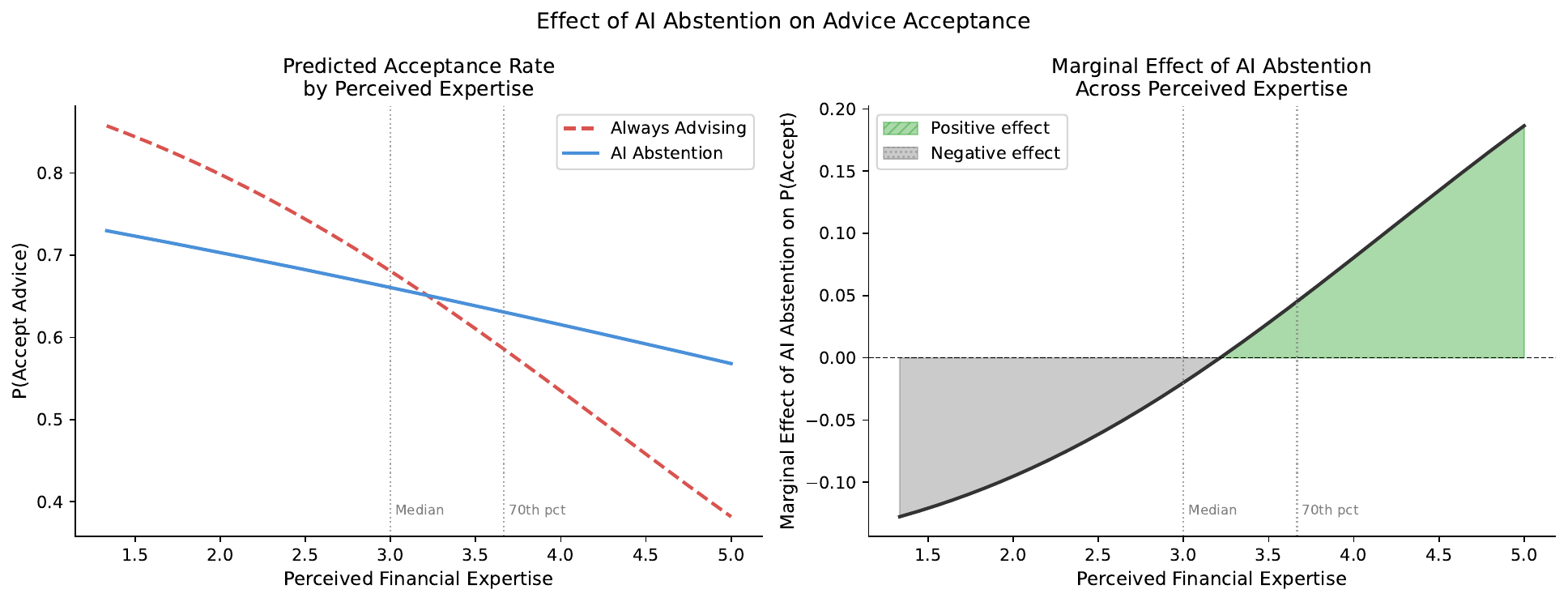}}
{Effect of AI Abstention on Advice Acceptance.\label{fig:predicted_marginal}}
{Left: predicted acceptance probability by condition across the range of perceived expertise, all controls held at their sample means. Right: marginal effect of AI abstention on acceptance probability; green shading indicates a positive effect, grey a negative one. Dashed lines mark the median and 70th percentile of perceived expertise.}
\end{figure}

Figure~\ref{fig:predicted_marginal} illustrates where the division falls more directly. The left panel displays predicted acceptance probabilities for both conditions across the full range of perceived expertise, showing that the two lines cross to the right of the median. The right panel plots the marginal effect of AI abstention, which is negative at low expertise levels and becomes positive as expertise increases, turning positive beyond the median. Together these results provide strong support for H1.

%\begin{figure}[t!]
%    \centering
%    \includegraphics[width=\textwidth]{Figures/fig2_treatment_by_continuous_expertise.pdf}
%    \caption{Left panel: predicted probability of accepting advice for always advising and AI abstention conditions across the full range of perceived financial expertise, with all controls held at their sample means. Right panel: marginal effect of AI abstention on acceptance probability across the expertise range. Red shading indicates regions where AI abstention reduces acceptance; green shading indicates regions where AI abstention increases acceptance. Grey dashed lines mark the median and 70th percentile of perceived expertise.}
%    \label{fig:predicted_marginal}
%\end{figure}

The moderation result is robust to alternative modeling choices. The Treatment $\times$ Perceived Expert interaction remains positive and significant when estimated via average marginal effects and a linear probability model, confirming the result is not an artifact of the logistic functional form, and when estimated using a binomial GLM across all post-intervention questions rather than just the two evaluation tasks. The interaction is also robust to controlling for baseline differences in advice receptivity across groups (see Appendices~\ref{app:functional_form} and \ref{app:baseline_acceptance}).

\subsection{The Role of Cognitive Burden}

\begin{figure}[t!]
\FIGURE
{\includegraphics[width=0.6\textwidth]{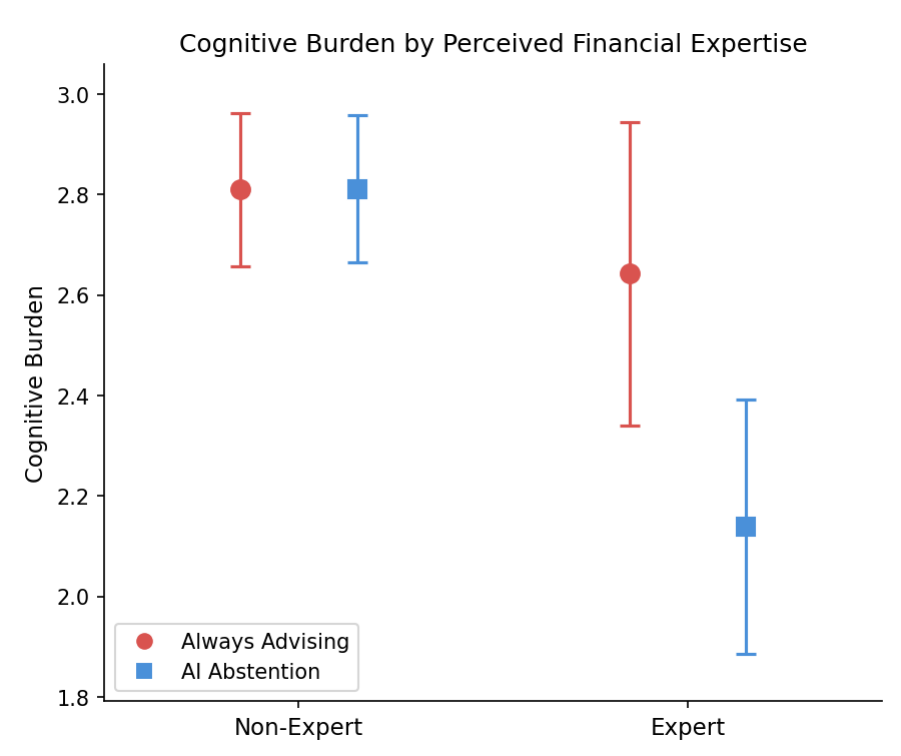}}
{Cognitive Burden by Perceived Financial Expertise.\label{fig:burden}}
{Mean cognitive burden (lower is less burden) by condition and perceived financial expertise group. Experts are defined as the top 30\% of the perceived expertise distribution (score $\geq 3.667$). Error bars represent 95\% confidence intervals.}
\end{figure}

Figure~\ref{fig:burden} displays mean cognitive burden experienced from reconciling AI advice by treatment condition and expertise group. AI abstention reduces cognitive burden for perceived expert users by $\approx20\%$ (2.642 in control vs.\ 2.139 in treatment, a reduction of 0.503 points on the 5-point scale) while having virtually no effect for non-expert users (2.810 in control vs.\ 2.811 in treatment). This asymmetric pattern, with burden reduced only for users with high perceived expertise, is consistent with H2.

%\begin{figure}[t!]
%    \centering
%    \includegraphics[width=0.8\textwidth]{Figures/fig3_burden.pdf}
%    \caption{Mean cognitive burden (lower is less burden) by condition and perceived financial expertise group. Experts are defined as the top 30\% of the perceived expertise distribution (score $\geq 3.667$). Error bars represent 95\% confidence intervals.}
%    \label{fig:burden}
%\end{figure}

To formally test H2, we estimate an OLS regression of the form:
\begin{equation}
    \text{Burden}_j = \alpha_0 + \alpha_1 \text{Treatment}_j + \alpha_2 \text{Expert}_j + \alpha_3 \text{Treatment}_j \times \text{Expert}_j + \mathbf{X}_j'\boldsymbol{\gamma} + \varepsilon_j
    \label{eq:h2}
\end{equation}

\noindent where $\text{Burden}_j$ is the perceived cognitive burden reported by participant $j$ in the post-survey; $\text{Treatment}_j$ is a binary indicator for assignment to the AI abstention condition; $\text{Expert}_j$ is a binary indicator for the top 30\% of the perceived expertise distribution; $\text{Treatment}_j \times \text{Expert}_j$ is their interaction; and $\mathbf{X}_j$ is a vector of person-level controls including two single-item pre-survey expertise measures, mean initial confidence, mean initial `correctness' (where `correct' indicates that the participant chose one of the two superior options), and the log of the mean initial (before AI advice) and final (after AI advice) response time averaged across the evaluation phase questions. All variables are at the person level ($N = 455$).

Table~\ref{tab:burden} reports the results. The Treatment $\times$ Perceived Expert interaction is negative and significant ($\beta = -0.518$, $p = 0.024$), indicating that AI abstention reduces cognitive burden significantly more for users with high perceived expertise than for non-expert users. The treatment main effect is near zero ($\beta = +0.017$, $p = 0.873$), suggesting that AI abstention has no meaningful effect on cognitive burden for non-expert users. These results provide formal support for H2, and this model serves as path $a$ in the moderated mediation analysis that follows.

\begin{table}[t!]
\TABLE
{OLS Regression: Effect of AI Abstention on Cognitive Burden\label{tab:burden}}
{%
\begin{tabular}{lcc}
\toprule
 & Model 1 & Model 2 \\
 & (Controls) & (No Controls) \\
\midrule
\textbf{Treatment (AI abstention)}              & 0.017     & 0.001 \\
                                                     & (0.108)   & (0.107) \\
\addlinespace
Perceived Expert (binary)                            & $-$0.094  & $-$0.168 \\
                                                     & (0.176)   & (0.170) \\
\addlinespace
\textbf{Treatment $\times$ Perceived Expert}         & \textbf{$-$0.518$^{**}$} & \textbf{$-$0.505$^{**}$} \\
                                                     & \textbf{(0.229)}         & \textbf{(0.226)} \\
\addlinespace
Controls                                             & Yes       & No \\
\addlinespace
\midrule
Observations                                         & 455       & 455 \\
$R^2$                                                & 0.061     & 0.049 \\
\bottomrule
\end{tabular}%
}%
{\NOTE{Notes.}OLS, HC3 robust SEs in parentheses. DV: post-survey cognitive burden (1–5 scale). Treatment coefficient reflects the effect of AI abstention for non-experts (reference group). Controls: two pre-survey expertise items, mean initial confidence, correctness, and log initial and evaluation-phase response time. $^{*}p < .10$, $^{**}p < .05$, $^{***}p < .01$.}
\end{table}

To examine whether the data are consistent with the proposed burden-reduction mechanism, we estimate the Average Causal Mediation Effect (ACME) of cognitive burden in the relationship between AI abstention and advice acceptance \citep{Imai2010}. The ACME decomposes the total treatment effect into an indirect component -- operating through changes in cognitive burden -- and a direct effect component. Following our theoretical model, we allow perceived expertise to moderate the effect of AI abstention on cognitive burden (path $a$), but leave the direct effect of AI abstention on acceptance unmoderated (path $c'$). Path $a$ is estimated via OLS at the person level; path $b$ is estimated via a question-level logistic regression with cognitive burden, treatment, perceived expertise, and the same controls as in Equation~\ref{eq:h1} as predictors of advice acceptance. Indirect effects are estimated using 2,000 bias-corrected accelerated (BCa) bootstrap \citep{Grun02102023} iterations at the person level, which handles the mixed model structure and accounts for any skewness in the distribution of indirect effect bootstrap samples.

\begin{table}[t!]
\TABLE
{Moderated Mediation Analysis: Cognitive Burden as Mediator\label{tab:mediation}}
{%
\begin{tabular}{lcc}
\toprule
 & Coefficient & 95\% CI \\
\midrule
\multicolumn{3}{l}{\textit{Path $a$: Treatment $\rightarrow$ Cognitive Burden (OLS, $N = 455$)}} \\
\addlinespace
Treatment (AI abstention)              & $+$0.017  & [$-$0.194, $+$0.229] \\
Perceived Expert (binary)              & $-$0.094  & [$-$0.439, $+$0.251] \\
Treatment $\times$ Perceived Expert    & $-$0.518  & [$-$0.966, $-$0.069] \\
\addlinespace
\midrule
\multicolumn{3}{l}{\textit{Path $b$ / Direct Effect ($c'$): Burden $\rightarrow$ Advice Acceptance (Logit, $N = 910$)}} \\
\addlinespace
Cognitive Burden                       & $-$0.221  & [$-$0.416, $-$0.026] \\
Treatment (AI abstention) ($c'$)       & $+$0.004  & [$-$0.380, $+$0.388] \\
Perceived Expert (binary)              & $-$0.659  & [$-$1.104, $-$0.214] \\
\addlinespace
\midrule
\multicolumn{3}{l}{\textit{Indirect Effects (BCa bootstrap, 2,000 iterations)}} \\
\addlinespace
ACME -- Expert users                   & $+$0.023  & [$+$0.003, $+$0.062] \\
ACME -- Non-expert users               & $-$0.001  & [$-$0.011, $+$0.008] \\
Index of Moderated Mediation (IMM)     & $+$0.023  & [$+$0.003, $+$0.065] \\
\addlinespace
\midrule
$R^2$ (path $a$)                       & 0.061 & \\
Pseudo $R^2$ (path $b$)                & 0.167 & \\
\bottomrule
\end{tabular}%
}%
{\NOTE{Notes.}Path-a controls: two pre-survey expertise items, mean initial confidence, correctness, and log initial and evaluation-phase response time. Path-b controls: initial correctness, confidence, log initial and final response time, question fixed effects, and view order.}
\end{table}

Table~\ref{tab:mediation} reports the results. Path $a$, the effect of AI abstention on cognitive burden, is significant and negative for perceived expert users (Treatment $\times$ Perceived Expert: $\beta = -0.518$, 95\% CI [$-0.966, -0.069$]), while the treatment main effect is near zero, suggesting that AI abstention has no meaningful effect on cognitive burden for non-expert users. Path $b$, the effect of cognitive burden on advice acceptance, is negative and significant ($\beta = -0.221$, 95\% CI [$-0.416, -0.026$]), indicating that higher cognitive burden reduces the probability of advice acceptance when controlling for AI abstention and perceived expertise. The direct effect of the user-facing AI abstention regime on acceptance, after accounting for the burden pathway, is small and non-significant ($\beta = +0.004$, 95\% CI $[-0.380,0.388]$). The ACME for perceived expert users is +0.023 (95\% CI [$+$0.003, $+$0.062]), indicating that AI abstention increases the probability of advice acceptance by approximately 2.3 percentage points through the cognitive burden pathway for perceived expert users. The ACME for non-expert users is essentially zero (95\% CI [$-$0.011, $+$0.008]), thus there is no evidence of the burden pathway for this group. The index of moderated mediation (IMM) is $+0.023$ (95\% CI [$+$0.003, $+$0.065]), confirming that the indirect effect differs significantly across expertise groups. This pattern replicates with an alternative estimation approach, Hayes Process Model 7 \citep{Hayes2015}, using a person-level binary outcome (see Appendix~\ref{app:mediation_robust}).

Together, these results support H3: cognitive burden is a pathway through which AI abstention increases advice acceptance among users with high perceived expertise. Consistent with our theoretical account, the indirect effect is present for perceived experts and absent for non-experts.

\subsection{Treatment Effect Dynamics and Appropriate Reliance}

\subsubsection{When Does the Expert Effect Emerge?}

\begin{figure}[t]
    \centering
    \includegraphics[width=0.7\textwidth]{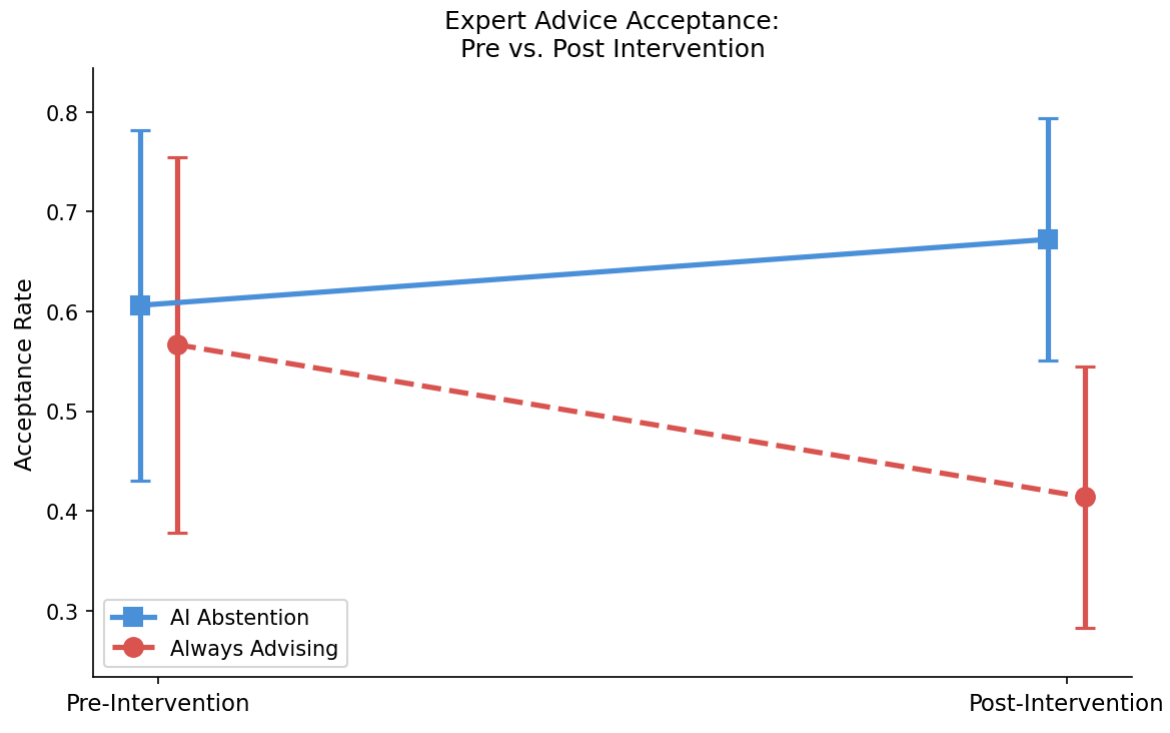}
    \caption{Advice acceptance rates for expert participants before and after experiencing the treatment manipulation, restricted to questions where the participant's initial answer was incorrect. Error bars represent 95\% confidence intervals.}
    \label{fig:decomposition}
\end{figure}

The crossover interaction documented above raises a natural question about the dynamics underlying the expert effect: does AI abstention actively increase expert acceptance, or does it prevent a decline that would otherwise occur? To examine this, we decompose the treatment effect for expert users into pre-intervention and post-intervention periods, where an intervention for a user is defined as the first question at which they were exposed to the randomly selected treatment. Because advice that recommends an alternative option in the pre-treatment phase is only provided when the participant's initial answer is not one of the two superior options, we restrict the sample to incorrect initial responses throughout both periods to ensure comparability across phases. Figure~\ref{fig:decomposition} and Table~\ref{tab:decomposition} report the results. Prior to experiencing any advice abstention, treatment and control experts accept at nearly identical rates (60.6\% vs.\ 56.7\%, $+$3.9 percentage points, $p = 0.784$). This also means that there is no evidence to support that the observed effect may have been caused by the instructions alone. Following the AI abstention experience, however, acceptance diverges sharply: treatment experts slightly increased (60.6\% to 67.2\%), while control experts declined markedly (56.7\% to 41.4\%), yielding a post-intervention difference of $+$25.8 percentage points ($p = 0.013$). The pre-to-post slope is marginally significant and negative for control experts ($\beta = -0.617$, $p = 0.082$) and essentially flat for treatment experts ($\beta = 0.287$, $p = 0.551$). These results suggest that AI abstention does not increase expert acceptance -- it preserves it, mitigating the progressively diminishing engagement with AI advice that develops when experts are exposed to continuous AI advice.

\begin{table}[t]
\TABLE
{Treatment Effect Decomposition: Expert Advice Acceptance (Incorrect Responses Only)\label{tab:decomposition}}
{%
\begin{tabular}{lcc}
\toprule
& Coef. & (SE) \\
\midrule
\multicolumn{3}{l}{\textit{Treatment vs.\ Control}} \\[4pt]
\quad Pre-Intervention & $0.163$ & $(0.592)$ \\
\quad Post-Intervention & $1.066^{**}$ & $(0.431)$ \\[6pt]
\multicolumn{3}{l}{\textit{Pre $\to$ Post slope}} \\[4pt]
\quad Control & $-0.617^{*}$ & $(0.355)$ \\
\quad Treatment & $0.287$ & $(0.481)$ \\
\bottomrule
\end{tabular}%
}%
{$^{*}p < .10$, $^{**}p < .05$, $^{***}p < .01$. Logit with clustered SEs by participant, estimated without controls given the small pre-intervention cell sizes. Sample restricted to expert participants on incorrect responses only ($N = 182$). Coefficients are log-odds.}
\end{table}

\subsubsection{Reliance Quality and Final Accuracy}

Because the evaluation tasks use a dual-optimal structure, accepting advice after an initially correct response is not harmful: the AI recommends an alternative option that is also appropriate. However, in such cases, accepting the advice would not improve accuracy. We therefore distinguish accuracy-improving acceptance, where the participant's initial answer was incorrect, from neutral acceptance, where the participant's initial answer was already appropriate. Table~\ref{tab:appropriate_reliance} reports the treatment effect on acceptance separately for each case by expertise group. For expert users, AI abstention produces a large and marginally significant increase in acceptance when their initial answer is incorrect ($\beta = +0.969$, $p = 0.055$), but no significant effect when their initial answer is correct ($\beta = +0.500$, $p = 0.214$). For non-expert users the pattern differs: AI abstention reduces acceptance in both cases. Taken together, these results suggest that the increase in expert acceptance under abstention is concentrated in the case where accepting advice is appropriate, rather than occurring indiscriminately across both cases.

Furthermore, this shift in acceptance patterns may also translate into a downstream improvement in decision quality. To test this, we re-estimate our primary H1 specification (Model 1) with final answer accuracy, rather than acceptance, as the outcome. The Treatment $\times$ Perceived Expert interaction is positive and significant ($\beta = +1.017$, $p = 0.036$; Table~\ref{tab:final_accuracy}), indicating that experts in the AI Abstention condition were more often correct on their final answer, not just more willing to accept advice.

\begin{table}[t]
\TABLE
{Treatment Effect on Advice Acceptance by Expertise and Initial Correctness (Q7--8)\label{tab:appropriate_reliance}}
{%
\begin{tabular}{llcc}
\toprule
& & \multicolumn{2}{c}{Treatment Effect (log-odds)} \\
\cmidrule(lr){3-4}
Group & Initial Answer & Coef. & (SE) \\
\midrule
\multirow{2}{*}{Experts}
  & Incorrect (beneficial advice) & $+0.969^{*}$ & $(0.506)$ \\
  & Correct (neutral advice) & $+0.500$ & $(0.402)$ \\
\midrule
\multirow{2}{*}{Non-Experts}
  & Incorrect (beneficial advice) & $-0.225$ & $(0.326)$ \\
  & Correct (neutral advice) & $-0.310$ & $(0.249)$ \\
\bottomrule
\end{tabular}%
}%
{$^{*}p < .10$, $^{**}p < .05$, $^{***}p < .01$. Logit with clustered SEs by participant, estimated without controls given the small subgroup cell sizes. Sample restricted to evaluation questions (Q7--8).}
\end{table}

\begin{table}[t]
\TABLE
{Treatment $\times$ Expertise Interaction, Final Answer Accuracy (Exploratory)\label{tab:final_accuracy}}
{%
\begin{tabular}{lcc}
\toprule
 & Coef. (log-odds) & (SE) \\
\midrule
TreatmentCondition $\times$ Perceived Expert & $+1.017^{**}$ & $(0.485)$ \\
\bottomrule
\end{tabular}%
}%
{$^{*}p < .10$, $^{**}p < .05$, $^{***}p < .01$. Outcome: final answer correct (post-advice). Logit with clustered SEs by participant. Full controls as in main H1 specification (Model 1). Sample restricted to evaluation questions (Q7--8), N=910.}
\end{table}

\subsection{Alternative Explanations}
\label{sec:alt_explanations_summary}
This section examines five alternative explanations for the crossover effect. We find no support for any of them; full analyses are reported in Appendix~\ref{appendix:altexplanations}.

\subsubsection{Perceived Expertise Versus Objective Accuracy}
A potential explanation is that perceived expertise proxies for objective accuracy. To test this, we replace perceived expertise with two measures of objective initial accuracy in the primary H1 specification. We find that neither measure interacts with treatment (all $p > 0.27$), suggesting that belief about one's own competence, not competence itself, is the operative moderator (Appendix~\ref{app:objective_accuracy}).

\subsubsection{Reduced Disagreement Exposure and Reactance}
A second possibility is that control experts, who by design encounter more AI disagreement during the initial phase, become reactive to this disagreement, and it is this reactance that drives their declining acceptance, rather than AI abstention itself improving receptivity. We test this three ways: whether disagreement frequency predicts evaluation-phase acceptance among control experts, whether the pre-to-post decline is impacted by disagreement exposure, and whether the expertise-moderated pattern is better explained by study phase or by cumulative disagreements experienced. We find that disagreement frequency does not predict subsequent acceptance ($\beta = -0.017$, $p = 0.949$), that the decline in advice acceptance is if anything larger among control experts exposed to \emph{fewer} disagreements, and that the expertise-moderated pattern is significant when indexed by phase ($\beta = 1.821$, $p = 0.022$) but not by cumulative disagreements ($\beta = 0.093$, $p = 0.768$) (Appendix~\ref{app:reactance}).

\subsubsection{Scarcity and Quality Signaling via Confidence in the AI}
A third possibility is that abstention makes the advice that is given seem scarcer and therefore more credible, inflating experts' confidence in the AI and thereby increasing acceptance. To test this, we examine whether treatment differentially raises expert users' confidence in the AI's recommendations. We find no evidence of this: the Treatment $\times$ Perceived Expert interaction on confidence in the AI is small and non-significant ($\beta = 2.773$, $p = 0.489$) (Appendix~\ref{sec:scarce_quality_sig}).

\subsubsection{Self-Confidence Updating}
A fourth possibility is that selective advice shifts users' confidence in their own judgment ahead of the evaluation phase. To test this, we examine whether treatment differentially affects initial self-confidence on the evaluation questions. We find no such effect ($\beta = -0.982$, $p = 0.833$) (Appendix~\ref{sec:conf_update}).

\subsubsection{Experimental Demand from Instruction Framing}
A final possibility is that the treatment instructions, which describe selective advising favorably, created a demand effect independent of the experienced abstention regime. To test this, we compare acceptance before participants experience any abstention, check whether the identical framing of AI abstention affects all users uniformly, and test for post-randomization differences in confidence or accuracy attributable to the instructions alone. We find three patterns inconsistent with a demand account: treatment and control experts accept advice at indistinguishable rates prior to the first experienced abstention ($p = 0.784$), the identical framing \emph{reduces} acceptance among non-experts rather than inflating it uniformly, and the instructions produce no post-randomization differences in confidence or accuracy (Appendix~\ref{app:demand}).

Together, these five checks point to the experienced abstention regime, rather than its description, its scarcity value, or objective competence, as the source of the crossover effect.

\section{Discussion}
AI systems that abstain from advising in some cases are rapidly becoming a practical reality, yet theories of advice taking have been built on the assumption that advice is always present. We studied what happens when that assumption is relaxed: participants made a series of financial decisions assisted either by an AI that abstained on some decisions or by one that always advised, with objective advice quality held constant across conditions. We find that AI abstention increases advice acceptance among users with high perceived expertise and tends to reduce it among users with low perceived expertise, that the expert-side effect is a preservation of receptivity to AI advice that would otherwise erode under continuous advising, and that this effect is accounted for by the relief of cognitive burden. Below we discuss the implications of these findings for theory and for practice.

\subsection{Theoretical Implications}

Our findings carry three implications for theories of human-AI collaboration and advice taking.
First, the AI abstention regime is a design property of AI advisory artifacts that must be theorized in its own right. Research on AI decision support has largely framed the design problem as one of advice quality: how accurate, explainable, trustworthy, or persuasive a recommendation should be. Correspondingly, advice-taking theories typically presupposes the presence of advice, treating the user's problem as one of weighting a recommendation that is always there \citep{SniezekVanSwol2001_OBHDP_TrustConfidenceAdvice, Yaniv2004_OBHDP_KnowledgeAdvice}. Our results establish a distinct locus of design. 
%Because advice quality was held constant across our conditions the effects we document are attributable to AI abstention itself, experienced as a user-facing regime. 
Whether the system advises on every decision or reserves the ability to abstain has behavioral consequences that are independent of, and irreducible to, the quality of the advice ultimately provided. Theories of reliance on AI must therefore incorporate when the system advises alongside what it recommends, and the growing technical literature on abstention mechanisms \citep{Wen2025_TACL_KnowYourLimits, Feng2024_ACL_DontHallucinate, fanconi2026cascaded} has a behavioral analog: every abstention policy is also a user experience whose effects require theorizing.

Second, the behavioral consequences of AI abstention are expertise-contingent. 
Prior work supports two opposing predictions. Work on user responses to abstention has examined immediate perceptions of safety-driven refusals and uncertainty expressions in factual information retrieval \citep{Wester2024, Kim2024_FAccT_UncertaintyExpression, zheng-etal-2025-easy}, finding that responses are largely unfavorable, which predicts that withholding advice depresses subsequent advice-taking. Work on the cognitive demands of continuous advising points the other way: because the effort of evaluating and reconciling recommendations drives disengagement over time \citep{lebovitz_engage_2022, ancker2017, You03042022}, withholding advice should relieve that demand and may sustain acceptance. Neither literature has examined downstream advice-taking behavior under abstention directly. Our results show that users respond to the AI abstention regime through the lens of their own perceived expertise, and thus both predictions are correct, depending on the user. For users with low perceived expertise, the absence of advice does not relieve the burden of processing advice, perhaps because it is a withdrawal of support they depend on, and receptivity does not see the same positive effect for low-expertise users as it does for high-expertise users, consistent with prior findings that abstention and refusal are unfavorably received \citep{Wester2024, zheng-etal-2025-easy}. For users with high perceived expertise, the same absence relieves the demand of evaluating yet another recommendation, and downstream receptivity increases. 
%The same regime, identically framed and communicated to all users, can produce opposite behavior for different types of users. 
This implies that the effect of AI abstention cannot be characterized by an average treatment effect, and that theories of reliance must consider user characteristics. It also contributes to the expertise-contingent reliance literature \citep{Logg2019_OBHDP_AlgorithmAppreciation, Jussupow2021_ISR_MedDiagnosisAI}: perceived expertise moderates responses to AI advisors, and our analyses show it is perceived rather than objective expertise that is operative, consistent with the argument that attitudes are formed from the subjective interaction experience.

Third, we identify cognitive burden as the mechanism linking the user-facing abstention regime to sustained engagement, which supplies an explanation for a pattern prior research has documented but left underexplained. Reliance on AI advice
is constrained not only by whether users trust a recommendation, but by the cognitive effort of deciding whether to rely on it
at all; that cost recurs with
every recommendation received. We find that AI abstention reduces the
cognitive burden reported by high-perceived-expertise users while leaving
non-experts' burden unchanged, and that this differential relief
statistically accounts for the differential effect on acceptance. Because burden is measured after the decision tasks, we interpret this as evidence consistent with a burden-reduction mechanism rather than a demonstration that
burden relief drives the effect. The explanatory value of this mechanism is visible in our control condition, which reproduces the progressive expert disengagement that prior studies have documented over extended professional engagements \citep{lebovitz_engage_2022, Jussupow2021_ISR_MedDiagnosisAI, ancker2017}: high-perceived-expertise users accepted progressively less advice even though the advice was always objectively appropriate. The erosion is driven by the accumulating cost of evaluating advice, not by advice failures---a distinction that existing studies focused on trust and advice quality cannot capture, and that positions cognitive burden as an under-examined outcome variable in research on AI advising system design.

Two refinements bound these claims. The expert effect is more precisely
a preservation than a gain: decomposing the effect over the course of the interaction shows that AI abstention does not raise expert acceptance above its starting level but prevents the decline that continuous advising produces. Additionally, an exploratory analysis shows that the effect of AI abstention on expert acceptance is larger when the user's initial answer is incorrect---when accepting would improve their decision---than when it is already correct.
This raises the possibility that AI abstention also positively affects which type of advice high-perceived-expertise users accept---a question of appropriate reliance \citep{PassiVorvoreanu2022_MSRTR_Overreliance, Nourani2020_HCOMP_DomainExpertiseTrust, Snyder2026} that we regard as a promising direction for future work.

\subsection{Practical Implications}

For organizations and designers deploying AI advisory systems, our findings
translate into four implications.

First, whether to include an abstention mechanism is a deployment decision
that must be made while considering the user population. Because the effect of AI
abstention splits in direction across users, a system that improves
receptivity in a population of high-perceived-expertise users may
simultaneously reduce it in a population of low-perceived-expertise users.
Average-level pilots and A/B tests will mask this: an abstaining system
evaluated on a mixed population may show a null average effect while helping
one segment and harming another. Deployment evaluations should therefore be
segmented by relevant user characteristics, such as perceived expertise. Where the population is
heterogeneous, no single provision policy suits everyone, and the abstention
mechanism, or its presence at all, should be calibrated to the user:
abstention-enabled configurations for confident, experienced users, and
consistent advising for users who seek and depend on continuous guidance.
Perceived expertise can be measured with short self-report scales of the
kind we used, making such calibration practically feasible at onboarding.

Second, for expert-facing deployments specifically, AI abstention offers a possible remedy to a persistent and costly problem in professional AI adoption: disengagement and abandonment. Radiologists disengage from AI inputs rather than expend the effort of reconciling them
\citep{lebovitz_engage_2022}, clinical decision support systems that generate high
volumes of alerts produce fatigue that suppresses compliance
\citep{ancker2017}, and elevated cognitive burden reduces adherence to
recommendations across settings \citep{Wogalter1999, Wogalter1998, You03042022}.
Our results identify why remedies focusing exclusively on quality of AI output have limited impact on this problem: the erosion we observe occurs even when every recommendation is
objectively appropriate, so improving advice quality alone cannot solve it. An
AI abstention regime intervention instead addresses the source of the
burden---the volume of recommendations that must be evaluated and
reconciled. Organizations observing declining engagement with an AI advisory tool among experienced users should consider that the system may be advising too often.

Third, the burden channel implies that advice volume is a cost to be managed,
not a feature to be maximized. Expert users in professional settings already
operate under substantial cognitive demand, and systems that compound it risk
undermining receptivity regardless of the quality of the advice they provide.
Designers commonly treat coverage---providing input on every case---as
unambiguously valuable; our findings indicate that for confident users 
additional recommendations carry an evaluative cost, and that suppressing
low-value recommendations is itself a form of user support. Notably, the
well-established technical benefit of abstention, improved average advice
quality, was deliberately switched off in our design; in real deployments the
behavioral benefit we document and the quality benefit would compound.

Fourth, how the advising policy is communicated is part of the design. Our
treatment framed abstention as a deliberate, beneficial policy, with
instructions validated through iterative pre-testing, and we regard such transparent framing as a baseline requirement: a system that
withholds advice without explanation invites users to read the silence as
failure, incompetence, or arbitrary withholding. Two features of our results
sharpen this guidance. The sentiment of the framing was identical across conditions
and across the expertise distribution, yet abstention still reduced
acceptance among users with low perceived expertise, so clear communication
alone does not neutralize the downside for guidance-dependent users,
reinforcing the case for calibration rather than messaging fixes. And because
our estimates reflect a well-communicated implementation, they are best read
as the effects of abstention done carefully; opaque or poorly explained
implementations may produce worse outcomes for all user groups.

\subsection{Limitations and Future Work}

Several limitations of this study should be noted. First, our experiment uses
financial decision tasks as an accessible analog for high-stakes uncertain
decision-making settings. While financial decisions share the key structural
properties of these settings such as genuine ambiguity, no immediate feedback, and
varying levels of user expertise, future work should examine whether the
patterns we document generalize to professional domains such as medical
diagnosis or legal judgment, where the stakes are higher and the expertise
distribution may be more compressed.

Second, our findings concern perceived financial expertise among online
participants making structured financial decisions, not credentialed
professionals making consequential decisions in the field. This design is
appropriate for testing our psychological argument, which centers on users'
perceived competence in the task domain. Importantly, credentialed
professionals in high-stakes settings tend to possess both high objective
expertise, meaning they are generally good at their tasks, and high perceived
expertise, meaning they are generally confident in their judgment. Our
findings show that it is perceived expertise that drives the AI abstention
effect, and perceived expertise is the dimension along which real-world
professionals are most clearly distinguished from novices. While future work
should examine this directly in professional settings, our results suggest
the mechanism identified here is likely to be operative among credentialed
experts making consequential decisions.

Third, our experiment exposes participants to a brief, approximately
fifteen-minute interaction with a short withholding experience. In practice,
AI abstention systems are used repeatedly over extended periods, and the
effects we document may evolve with experience. Notably, even within this
brief interaction we observed the progressive decline in expert receptivity
under continuous advising that prior work has documented in extended
professional engagements \citep{lebovitz_engage_2022, ancker2017}, lending credence
to the ecological validity of our design. Nevertheless, it remains important
to understand whether the benefits of AI abstention persist and 
accumulate over longer engagements, and longitudinal research in professional
settings is a natural next step.

Finally, our study examines a single operationalization of AI abstention,
framed and communicated to the user as a mechanism that provides advice when
its input is judged to be beneficial. This framing reflects what we believe
to be best practice, and our instructions were developed and validated
through iterative pre-testing of the kind a field deployment should involve.
Our theoretical account suggests why this operationalization matters and how
to move beyond it: because users interpret abstention through the policy
rationale communicated to them, alternative framings such as expressing
uncertainty, offering no explanation, or attributing abstention to different rationales, are not merely alternative treatments but alternative regimes,
each of which may be read differently across the expertise distribution.
Understanding how the communicated content of the advice-provision policy
shapes user responses, and whether the cognitive burden mechanism identified
here generalizes across regimes, is a theory-driven agenda for future work.
Our study establishes the responses that well-designed, clearly communicated
AI abstention can produce; mapping the space of less favorable
operationalizations remains open.

\section{Conclusion}
AI advising systems are being deployed at scale in consequential domains, and substantial research attention has been devoted to when and how they should advise. Yet how users respond when those systems choose not to advise has received far less attention. In this paper, we show that the absence of advice has real and heterogeneous effects on how users subsequently engage with AI recommendations, and that those effects cannot be anticipated from the average effect. For users who perceive themselves to be experts in the task domain, an AI that advises selectively preserves their willingness to engage over time, relieving the cognitive toll of repeatedly reconciling their judgment with a system that always speaks. For users who do not perceive themselves to be experts and who depend more heavily on guidance, that same absence erodes their engagement. The design of when to advise is therefore not a secondary implementation detail but a first-order property of the advisory artifact, and theories of human-AI collaboration must account for the fact that the same silence speaks differently to different users. Realizing the promise of AI abstention depends on calibrating it to the
people the system is meant to serve.

%\newpage

\bibliographystyle{informs2014}  % Changed from INFORMS
%\bibliography{references}
{\small \bibliography{references}}

\newpage
\begin{APPENDICES}

\section{Robustness Checks}
\label{appendix:robustness}

\subsection{Post-Exclusion Randomization Checks}
\label{app:randomization}

Table~\ref{tab:randomization} reports pre-treatment characteristics of the final analytic sample by condition. Treatment and control participants do not differ significantly on perceived financial expertise ($t = 0.100$, $p = 0.921$), initial confidence ($t = -0.056$, $p = 0.955$), or initial accuracy ($t = 0.427$, $p = 0.670$), confirming successful randomization, and providing further evidence that the instructions alone do not lead to any meaningful difference in behavior on tasks immediately following the instructions.

Participants were excluded if they failed a condition-specific comprehension check assessing understanding of the advising regime they were assigned to, failed an attention check, or were flagged by automated bot detection, all of which were pre-registered exclusion criteria. The exclusion criterion was applied identically across conditions -- each participant was assessed on their own condition's instructions. The analytic sample remains balanced across conditions on all pre-treatment characteristics after exclusions (Table~\ref{tab:randomization}).

\begin{table}[h!]
\TABLE
{Randomization Check: Pre-Treatment Characteristics by Condition\label{tab:randomization}}
{%
\begin{tabular}{lcccc}
\toprule
 & Control & Treatment & $t$ & $p$ \\
 & $M$ (SD) & $M$ (SD) & & \\
\midrule
Perceived Financial Expertise & 3.175 (0.890) & 3.167 (0.945) & 0.100 & 0.921 \\
Initial Confidence (Q1--2)    & 60.751 (23.939) & 60.878 (24.317) & $-$0.056 & 0.955 \\
Initial Accuracy (Q1--2)      & 0.761 (0.306) & 0.748 (0.323) & 0.427 & 0.670 \\
\midrule
$N$ & 213 & 242 & & \\
\bottomrule
\end{tabular}%
}%
{\NOTE{Notes.} Two-sample $t$-tests. No significant differences between conditions on any pre-treatment characteristic.}
\end{table}

\subsection{Functional Form and Outcome Scope}
\label{app:functional_form}

We verify that the crossover interaction is not an artifact of the logistic functional form or the restriction of the primary analysis to evaluation questions Q7 and Q8. Table~\ref{tab:robustness_functional} reports three additional specifications. First, we compute average marginal effects from the primary logit specification. The marginal effect of the Treatment $\times$ Perceived Expert interaction is $+0.190$ ($p = 0.014$), indicating that the effect of AI abstention on advice acceptance is approximately 19 percentage points more positive for perceived expert users than for non-expert users. Second, we re-estimate the primary model using a linear probability model with participant-clustered standard errors. The Treatment $\times$ Perceived Expert interaction coefficient on the probability scale remains positive, similar to the average marginal effect from the primary logit specification, and significant ($\beta = +0.206$, $p = 0.015$), confirming the result is not an artifact of the logistic functional form. Third, we estimate a binomial GLM using a broader set of post-intervention questions rather than just Q7 and Q8, with each participant contributing the number of advice instances accepted out of all eligible post-intervention instances. The Treatment $\times$ Perceived Expert interaction is positive and significant across all four binomial model specifications ($p = 0.010$--$0.032$), with somewhat stronger significance than the primary logit models. Together, these checks confirm that the moderation result is robust to both the choice of functional form and the scope of the outcome measure.

\begin{table}[t!]
\TABLE
{Robustness Checks: Functional Form\label{tab:robustness_functional}}
{%
\begin{tabular}{lccc}
\toprule
 & AME & LPM & \\
 & (1) & (2) & \\
\midrule
Treatment (AI Abstention) & $-$0.048 & $-$0.043 & \\
                               & (0.041)  & (0.040)  & \\
\addlinespace
Perceived Expert (binary)      & $-$0.202 & $-$0.230 & \\
                               & (0.055)  & (0.063)  & \\
\addlinespace
\textbf{Treatment $\times$ Perceived Expert} & \textbf{$+$0.190$^{**}$} & \textbf{$+$0.206$^{**}$} & \\
                                             & \textbf{(0.077)}         & \textbf{(0.085)}         & \\
\addlinespace
Controls & Yes & Yes & \\
\midrule
Observations & 910 & 910 & \\
\bottomrule
\end{tabular}%
}%
{\NOTE{Notes.} Model 1 reports average marginal effects from the primary logit specification. Model 2 reports a linear probability model with participant-clustered standard errors. All coefficients are on the probability scale. Controls include initial correctness, initial confidence, log initial and final response time, question fixed effects, and view order. $^{*}p < .10$, $^{**}p < .05$, $^{***}p < .01$.}
\end{table}

\begin{table}[t!]
\TABLE
{Binomial GLM Robustness: Moderation of AI Abstention Effect by Perceived Expertise\label{tab:binomial}}
{%
\begin{tabular}{lcccc}
\toprule
 & Model 1 & Model 2 & Model 3 & Model 4 \\
\midrule
\textbf{Treatment (AI Abstention)}                  & $-$1.665$^{**}$ & $-$1.817$^{**}$ & $-$0.287  & $-$0.381$^{*}$ \\
                                                         & (0.820)         & (0.822)         & (0.215)   & (0.219) \\
\addlinespace
Perceived Expertise (cont.)                              & $-$0.990$^{***}$ & $-$0.654$^{***}$ &          & \\
                                                         & (0.187)          & (0.189)          &          & \\
\addlinespace
Perceived Expert (binary)                                &                  &                  & $-$1.579$^{***}$ & $-$1.049$^{***}$ \\
                                                         &                  &                  & (0.291)          & (0.312) \\
\addlinespace
\textbf{Treatment $\times$ Perceived Expertise (cont.)}  & \textbf{0.502$^{**}$} & \textbf{0.527$^{**}$} &          & \\
                                                         & \textbf{(0.234)}      & \textbf{(0.236)}      &          & \\
\addlinespace
\textbf{Treatment $\times$ Perceived Expert (binary)}    &                  &                  & \textbf{1.005$^{***}$} & \textbf{1.039$^{**}$} \\
                                                         &                  &                  & \textbf{(0.393)}       & \textbf{(0.408)} \\
\addlinespace
Controls                                                 & No               & Yes              & No        & Yes \\
\midrule
Observations                                             & 455              & 455              & 455       & 455 \\
\bottomrule
\end{tabular}%
}%
{\NOTE{Notes.} Coefficients are from binomial GLM with HC3 robust standard errors in parentheses. Outcome is the number of advice instances accepted out of total eligible post-intervention instances ($N = 455$ participants). Models 1 and 2 use perceived expertise as a continuous moderator; Models 3 and 4 use the binary perceived expert indicator. Controls include mean initial confidence and log mean initial response time. $^{*}p < .10$, $^{**}p < .05$, $^{***}p < .01$.}
\end{table}

\subsection{Alternative Expertise Cutoffs}
\label{app:cutoffs}

Our primary specification classifies participants as perceived experts if their perceived financial expertise score exceeds 3.667, the 70th percentile of the distribution. To verify that the moderation result does not depend on this particular threshold, we re-estimate the primary H1 model (Model 1) at distinct cutpoints on the expertise scale, retaining those that leave at least 20 participants in the expert group. Because the expertise measure averages three items on a five-point scale, the distribution contains substantial ties, yielding five distinct thresholds.

Table~\ref{tab:cutoffs} reports the Treatment $\times$ Perceived Expert interaction at each. The interaction is positive and significant at every threshold at or above 3.667, and it increases in magnitude as the threshold rises, from $+$1.047 at the primary cutoff to $+$2.074 among the most confident participants. Notably, significance is retained as the threshold rises even though the expert group shrinks from 230 question-level observations to 60, indicating that the growth in effect size outpaces the accompanying loss of precision. At the two lowest thresholds the interaction is positive but not significant, consistent with our theoretical account: the effect emerges only once participants regard themselves as genuinely expert, rather than accumulating steadily across the expertise scale. This nonlinearity also explains why the continuous-moderator specifications in Table~\ref{tab:h1} are somewhat weaker than their binary counterparts, since a linear-in-expertise term averages over a relationship that is flat at lower levels and steep at higher ones. We note that these thresholds are nested within the same sample and therefore do not constitute independent tests; the relevant evidence is the monotone pattern rather than the individual coefficients.

\begin{table}[t!]
\TABLE
{Robustness: Treatment $\times$ Perceived Expert Interaction at Alternative Cutoffs\label{tab:cutoffs}}
{%
\begin{tabular}{lccc}
\toprule
Expertise cutoff & Expert obs. & Coef. & (SE) \\
\midrule
$> 3.000$          & 438 & $+0.291$          & $(0.387)$ \\
$> 3.333$          & 348 & $+0.625$          & $(0.390)$ \\
$> 3.667$ (primary) & 230 & $+1.047^{**}$     & $(0.436)$ \\
$> 4.000$          & 104 & $+1.312^{**}$     & $(0.650)$ \\
$> 4.333$          & 60  & $+2.074^{**}$     & $(0.949)$ \\
\bottomrule
\end{tabular}%
}%
{\NOTE{Notes.} Log-odds from logistic regression, participant-clustered SEs in parentheses. Each row re-estimates the primary H1 specification (Table~\ref{tab:h1}, Model 1) with the perceived expert indicator defined at the stated cutoff. Controls, question fixed effects, and sample ($N = 910$) are identical across rows. $^{*}p < .10$, $^{**}p < .05$, $^{***}p < .01$.}
\end{table}

\subsection{Controlling for Baseline Acceptance}
\label{app:baseline_acceptance}

To account for pre-existing differences in receptivity to AI advice, we re-estimate the primary H1 model adding pre-intervention advice acceptance as an additional control. Pre-intervention acceptance is computed among participants who answered at least one question incorrectly during the initial phase, yielding a sample of 228 participants contributing 456 observations. The Treatment $\times$ Perceived Expert interaction remains positive and significant ($\beta = +1.673$, $p = 0.027$), and pre-intervention acceptance is itself a strong positive predictor of evaluation-phase acceptance ($\beta = +2.299$, $p = 0.000$), as expected. The main finding is therefore not driven by baseline differences in advice receptivity. Full results are reported in Table~\ref{tab:baseline_acceptance}.

\begin{table}[t!]
\TABLE
{Robustness Check: Controlling for Pre-Intervention Acceptance\label{tab:baseline_acceptance}}
{%
\begin{tabular}{lccc}
\toprule
 & Coef. & SE & $p$ \\
\midrule
Treatment (AI Abstention)               & $-$0.794 & 0.340 & 0.020 \\
Perceived Expert (binary)                    & $-$1.923 & 0.588 & 0.001 \\
\textbf{Treatment $\times$ Perceived Expert} & \textbf{$+$1.673} & \textbf{0.756} & \textbf{0.027} \\
Pre-intervention acceptance                  & $+$2.299 & 0.342 & 0.000 \\
\midrule
Observations & \multicolumn{3}{c}{456} \\
\bottomrule
\end{tabular}%
}%
{\NOTE{Notes.} Logit with participant-clustered standard errors. Sample restricted to participants with at least one incorrect initial response prior to their first intervention (228 participants, 456 observations). Controls include initial correctness, initial confidence, log initial and final response time, question fixed effects, and view order. $^{*}p < .10$, $^{**}p < .05$, $^{***}p < .01$.}
\end{table}

\subsection{Mediation Robustness}
\label{app:mediation_robust}
To confirm the robustness of the H3 result, we estimate an alternative specification using Hayes Process Model 7 \citep{hayes2022introduction} as implemented in \emph{pyprocessmacro} \citep{andre2021pyprocessmacro}, with a person-level binary outcome indicating whether the participant accepted advice on both evaluation questions. Model 7 allows perceived expertise to moderate the effect of AI abstention on cognitive burden (path $a$) while leaving the direct effect unmoderated, consistent with the primary specification. Table~\ref{tab:mediation_robust} reports the index of moderated mediation (IMM $= +0.118$, 95\% CI [$+$0.009, $+$0.347]), confirming that the indirect effect of AI abstention on advice acceptance through cognitive burden is significantly larger for perceived expert users than for non-expert users. These results provide evidence that the moderated mediation finding is robust to the choice of outcome operationalization and choice of mediation model.

\begin{table}[h!]
\TABLE
{Mediation Robustness: Index of Moderated Mediation (Person-Level Binary Outcome)\label{tab:mediation_robust}}
{%
\begin{tabular}{lc}
\toprule
 & Coefficient \\
\midrule
\multicolumn{2}{l}{\textit{Conditional Indirect Effects}} \\
\addlinespace
ACME -- Non-expert users  & $-$0.004 \\
ACME -- Expert users      & $+$0.114 \\
\addlinespace
\midrule
Index of Moderated Mediation (IMM) & $+$0.118 \\
95\% CI                            & [$+$0.009, $+$0.347] \\
\bottomrule
\end{tabular}%
}%
{\NOTE{Notes.} Estimated using Hayes Process Model 7 \citep{hayes2022introduction} as implemented in pyprocessmacro, with a person-level binary outcome indicating whether the participant accepted advice on both evaluation questions. Perceived expertise moderates path $a$ only; the direct effect is unmoderated. Indirect effects are the product of path $a$ and path $b$ coefficients. Controls include the two evaluation-question indicators, mean initial confidence, mean initial correctness, and log mean initial and evaluation-phase response time. Confidence intervals based on 5,000 bias-corrected bootstrap samples.}
\end{table}

\section{Alternative Explanations}
\label{appendix:altexplanations}

This section examines alternative explanations that could plausibly lead to the main effects we observe. 

\subsection{Perceived Expertise Versus Objective Accuracy}
\label{app:objective_accuracy}

A potential alternative explanation for the moderating role of perceived expertise is that it merely proxies for objective accuracy.  We argued theoretically that in no-feedback settings, advice acceptance is driven not by actual accuracy -- which is not observable -- but by how competent decision-makers believe themselves to be. To test this empirically, we estimate logistic regression models replacing the perceived expertise moderator with two measures of objective accuracy: first-half accuracy (averaged across all initial phase questions) and Q1--2 accuracy (averaged across the first two questions only), each with and without controls. Table~\ref{tab:conf_acc} reports the Treatment $\times$ Accuracy interaction term from each model. None of the interactions approach significance (all $p > 0.27$), and the coefficients are small and inconsistent in sign across specifications. This stands in contrast to the significant and consistent Treatment $\times$ Perceived Expertise interaction documented in Table~\ref{tab:h1}. Objective accuracy does not seem to moderate the AI abstention effect while evidence supports that perceived expertise is the operative moderator. 

\begin{table}[t!]
\TABLE
{Logistic Regression: Treatment $\times$ Objective Accuracy Interactions\label{tab:conf_acc}}
{%
\begin{tabular}{lcccc}
\toprule
 & Model A & Model B & Model C & Model D \\
 & (No Controls) & (Controls) & (No Controls) & (Controls) \\
\midrule
 & \multicolumn{2}{c}{First-Half Accuracy} & \multicolumn{2}{c}{Q1--2 Accuracy} \\
\cmidrule(lr){2-3} \cmidrule(lr){4-5}
\addlinespace
\textbf{Treatment $\times$ Accuracy}   & $-$0.324  & $-$0.195  & 0.479   & 0.706 \\
                                       & (0.988)   & (0.951)   & (0.615) & (0.641) \\
                                       & $p=0.743$ & $p=0.837$ & $p=0.436$ & $p=0.270$ \\
\addlinespace
Controls                               & No        & Yes       & No      & Yes \\
Question fixed effects                 & No        & Yes       & No      & Yes \\
\addlinespace
\midrule
Observations                           & 910       & 910       & 910     & 910 \\
\bottomrule
\end{tabular}%
}%
{\NOTE{Notes.} Coefficients are log-odds from logistic regression reporting the Treatment $\times$ Accuracy interaction term only. Standard errors clustered at the participant level in parentheses. None of the interactions approach significance (all $p > 0.27$), confirming that objective accuracy does not moderate the AI abstention effect. Controls include initial correctness, initial confidence, log initial and final response time, and view order. $^{*}p < .10$, $^{**}p < .05$, $^{***}p < .01$.}
\end{table}

\subsection{Reduced Disagreement Exposure and Reactance}
\label{app:reactance}

\subsection{Reduced Disagreement Exposure and Reactance}
\label{app:reactance}

A second alternative interpretation of the expert effect is that control experts become annoyed, or reactive from repeated AI disagreement during the initial phase, lowering their subsequent acceptance, rather than AI abstention itself improving receptivity. Control participants experienced on average 3.05 disagreements during the initial phase ($SD = 0.97$), compared to 1.02 in the treatment condition ($SD = 0.92$), a difference that is by design.

To examine this, we test whether variation in disagreement frequency within the control condition predicts evaluation-phase acceptance for expert users. If reactance from disagreement drives the decline in control expert acceptance, then control experts who experienced more disagreements should show lower subsequent acceptance. Table~\ref{tab:disagreements} reports the results. Among control experts ($n = 55$), disagreement frequency does not significantly predict evaluation-phase acceptance ($\beta = -0.017$, $p = 0.949$). Among non-expert control participants, by contrast, disagreements are positively associated with acceptance ($\beta = 0.436$, $p = 0.037$). When disagreement frequency is included as a covariate in the primary H1 model, the Treatment $\times$ Perceived Expert interaction remains positive and significant ($\beta = 1.011$, $p = 0.022$).

We extend this analysis in two additional ways. First, we construct a question-level measure of cumulative disagreements experienced by each participant up to each question, and examine whether the pre-to-post decline in control expert acceptance varies by disagreement accumulation. Among control experts in the post phase, those who had experienced fewer cumulative disagreements (at or below the median of three) showed a decline from 56.7\% to 38.2\% acceptance (18.5 percentage points), while those who had experienced more disagreements declined from 56.7\% to 45.8\% (10.9 percentage points). The group with fewer disagreements showed a larger decline, contrary to what a reactance account would predict. A formal test finds that high versus low disagreement exposure does not significantly predict post-phase acceptance among control experts ($\beta = 0.313$, $p = 0.599$), and the pre-to-post decline remains directionally negative within the low-disagreement subgroup ($\beta = -0.748$, $p = 0.155$), suggesting the decline is associated with study phase rather than disagreement accumulation.

Second, we estimate two parallel three-way interaction models on all participants with incorrect initial responses in the pre and post phases ($N = 814$ observations, 374 participants), using the binary perceived expertise moderator consistent with the primary analysis. The first model tests whether the treatment effect on the pre-to-post slope is moderated by perceived expertise (Treatment $\times$ Post $\times$ Perceived Expert). The second replaces study phase with cumulative disagreements (Treatment $\times$ Cumulative Disagreements $\times$ Perceived Expert). In the first model, the three-way interaction is significant ($\beta = 1.821$, $p = 0.022$), with higher perceived expertise predicting a steeper pre-to-post decline under always-advising ($\beta = -1.161$, $p = 0.024$) and AI Abstention attenuating this decline. In the second model, the analogous three-way interaction is far from significant ($\beta = 0.093$, $p = 0.768$), while the main effect of cumulative disagreements is positive and significant ($\beta = 0.207$, $p = 0.024$), indicating that more disagreements is associated with higher rather than lower acceptance. Study phase moderated by expertise produces a significant pattern; disagreements moderated by expertise does not. Using continuous perceived expertise as the moderator produces the same conclusion (Treatment $\times$ Post $\times$ Perceived Expertise: $\beta = 0.984$, $p = 0.027$; Treatment $\times$ Cumulative Disagreements $\times$ Perceived Expertise: $\beta = 0.078$, $p = 0.631$).

Taken together, these analyses find no evidence within our data that variation in disagreement exposure accounts for the decline in expert acceptance under always-advising. The decline is present among control experts who experienced relatively few disagreements, more disagreements are associated with higher rather than lower acceptance, and the expertise-moderated pattern is significant when indexed by study phase but not when indexed by disagreement accumulation.
%We acknowledge that the design cannot fully disentangle abstention from reduced disagreement exposure and note this as a limitation, calling for future work with an additional control condition.

\begin{table}[t!]
\TABLE
{Disagreement Frequency and Evaluation-Phase Acceptance\label{tab:disagreements}}
{%
\begin{tabular}{lcccc}
\toprule
 & Coef. & SE & $p$ & $N$ \\
\midrule
\multicolumn{5}{l}{\textit{Effect of disagreement frequency on acceptance, within control condition}} \\
\addlinespace
Expert users & $-$0.017 & 0.263 & 0.949 & 55 \\
Non-expert users & $+$0.436 & 0.209 & 0.037 & 158 \\
\addlinespace
\midrule
\multicolumn{5}{l}{\textit{Primary H1 model controlling for disagreement frequency}} \\
\addlinespace
Treatment (AI Abstention) & $+$0.285 & 0.317 & 0.369 & \\
Perceived Expert (binary) & $-$1.100 & 0.322 & 0.001 & \\
Treatment $\times$ Perceived Expert & $+$1.011 & 0.441 & 0.022 & \\
First-half disagreements & $+$0.266 & 0.102 & 0.009 & \\
\addlinespace
\midrule
\multicolumn{5}{l}{\textit{Pre-to-post slope within control experts, by disagreement level}} \\
\addlinespace
Low cumulative disagreements ($\leq 3$) & $-$0.748 & 0.526 & 0.155 & 34 \\
High cumulative disagreements ($> 3$) & $-$0.435 & 0.379 & 0.251 & 24 \\
\addlinespace
\midrule
\multicolumn{5}{l}{\textit{Three-way interaction models (pre and post phase, incorrect only, $N = 814$)}} \\
\addlinespace
Treatment $\times$ Post $\times$ Perceived Expert & $+$1.821 & 0.797 & 0.022 & \\
Treatment $\times$ Cum.\ Disagreements $\times$ Perceived Expert & $+$0.093 & 0.313 & 0.768 & \\
\addlinespace
\midrule
Observations & \multicolumn{4}{c}{910 / 814} \\
\bottomrule
\end{tabular}%
}%
{\NOTE{Notes.} Top panel reports logistic regression coefficients for the effect of initial-phase disagreement frequency on evaluation-phase acceptance, estimated separately within the control condition. Second panel adds disagreement frequency as a covariate to the primary H1 logistic regression ($N = 910$). Third panel reports pre-to-post slopes within control experts separately for those below and above the median cumulative disagreement count (3), estimated on incorrect responses only. Bottom panel reports the key three-way interaction coefficients from two separate models estimated on all participants with incorrect initial responses in the pre and post phases ($N = 814$); the first indexes study progression by phase (pre vs.\ post), the second by cumulative disagreements experienced up to each question. The top two panels use the full control set from the primary specification. The bottom two panels exclude question fixed effects and view order, which are collinear with study phase; the third panel is estimated without controls given subgroup sizes of 24--34 participants. All coefficients are log-odds. Standard errors clustered at the participant level.}
\end{table}

\subsection{Scarcity and Quality Signaling via Confidence in the AI}
\label{sec:scarce_quality_sig}

A third alternative is that abstaining from providing advice makes later advice seem more valuable or credible because it is rarer and framed as high-confidence input, inflating participants' confidence in the AI's recommendations and thereby increasing acceptance. Confidence in the AI's recommendation is a well-documented predictor of advice acceptance \citep{WangLuYin2022_WWW_AcceptAIRecommendation, SniezekVanSwol2001_OBHDP_TrustConfidenceAdvice}. To test this potential scarcity effect, we examine whether AI abstention differentially raises confidence in the AI for expert users.

The Treatment $\times$ Perceived Expert interaction on confidence in the AI is small and non-significant ($\beta = 2.773$, $p = 0.489$), and the treatment main effect is also non-significant ($\beta = -3.576$, $p = 0.112$). We do not find enough evidence that AI abstention differentially raises expert confidence in the AI's recommendations. Confidence in the AI does strongly predict acceptance ($\beta = 0.021$, $p = 0.000$), consistent with prior work, but this reflects its general role as a predictor of advice utilization rather than evidence that it operates as a mechanism through which AI abstention differentially affects experts.

\subsection{Self-Confidence Updating}
\label{sec:conf_update}

A fourth alternative is that AI abstention produces selective advice that shifts users' confidence in their own initial judgments before the evaluation phase, and that this self-confidence shift drives the expert effect. We test whether treatment differentially affects initial confidence on the evaluation questions. The Treatment $\times$ Perceived Expert interaction on initial confidence is non-significant ($\beta = -0.982$, $p = 0.833$), and the treatment main effect is also non-significant ($\beta = -0.818$, $p = 0.760$). We find no evidence that AI abstention changes how confident participants are in their own answers prior to receiving AI advice.

\subsection{Experimental Demand from Instruction Framing} 
\label{app:demand}
A final alternative is that the treatment instructions, which describe selective advising favorably and characterize provided advice as high-confidence, high-value input, created an experimental demand to accept the advice that was offered, such that the effect reflects the description of the regime rather than the experience of it. Three observations argue against this account. First, the framing was delivered before any decision task, yet treatment and control experts accepted advice at statistically indistinguishable rates prior to the first experienced abstention ($+$3.9 percentage points, $p = 0.784$; Section 4.5.1, Table~\ref{tab:decomposition}): a demand effect operating through instructions should be present immediately, not emerge only after abstention is experienced. Second, the favorable framing was identical for all treatment participants, yet AI abstention reduced acceptance among users with low perceived expertise; instructions that inflate acceptance wholesale cannot generate a crossover. Third, the instructions produced no post-randomization differences in initial confidence, initial accuracy, or confidence in the AI (Appendix~\ref{app:randomization}; Sections~\ref{sec:scarce_quality_sig}--\ref{sec:conf_update}), as would be expected if participants had adopted a general belief in the system's superiority. We cannot rule out a demand effect that is conditional on both perceived expertise and the experience of abstention; but an account of that form is no longer a confound, it converges on our theoretical claim that the behavioral response is produced by the experienced regime rather than by its description.

\section{Participant Recruitment}
\label{appendix:recruitment}

\begin{table}[htbp]
\begin{rotate}

\TABLE
{Descriptive Statistics by Group, Phase, and Condition\label{tab:descriptive}}
{%
\setlength{\tabcolsep}{4pt}
\renewcommand{\arraystretch}{1.25}
\begin{tabular}{lllrr rrrrrrr}
\toprule
\textbf{Group} & \textbf{Phase} & \textbf{Cond.} & $N$ & $Q$
  & \textbf{Exp.} & \textbf{Init. Acc.} & \textbf{Final Acc.} & \textbf{Init. Conf.} & \textbf{AI Conf.} & \textbf{Init. Time} & \textbf{Final Time} \\
\midrule

% ALL
\multirow{6}{*}{\textbf{All}}
  & \multirow{2}{*}{Pre (Q1--2)}
    & T & 242 & 484  & 3.17 (0.94) & 0.75 (0.43) & 0.92 (0.28) & 60.88 (26.79) & 78.01 (21.74) & 62.52 (58.05) & 49.49 (47.45) \\
  & & C & 213 & 426  & 3.18 (0.89) & 0.76 (0.43) & 0.91 (0.29) & 60.75 (26.34) & 78.86 (20.10) & 61.31 (65.13) & 50.53 (57.28) \\
\cmidrule(l){2-12}
  & \multirow{2}{*}{During (Q3--6)}
    & T & 242 & 968  & 3.17 (0.94) & 0.73 (0.44) & 0.83 (0.37) & 56.16 (27.81) & 75.91 (21.33) & 51.94 (45.27) & 23.15 (29.84) \\
  & & C & 213 & 852  & 3.18 (0.89) & 0.72 (0.45) & 0.91 (0.29) & 59.64 (27.15) & 75.86 (21.54) & 51.63 (47.92) & 43.04 (44.61) \\
\cmidrule(l){2-12}
  & \multirow{2}{*}{Eval (Q7--8)}
    & T & 242 & 484  & 3.17 (0.94) & 0.62 (0.48) & 0.86 (0.35) & 52.29 (30.68) & 68.11 (25.90) & 48.82 (43.49) & 43.09 (32.88) \\
  & & C & 213 & 426  & 3.18 (0.89) & 0.62 (0.49) & 0.87 (0.34) & 53.11 (31.36) & 71.30 (23.55) & 47.40 (47.73) & 48.23 (44.85) \\

\midrule

% EXPERTS
\multirow{6}{*}{\textbf{Experts}}
  & \multirow{2}{*}{Pre (Q1--2)}
    & T & 60  & 120  & 4.37 (0.44) & 0.78 (0.41) & 0.88 (0.32) & 73.75 (21.23) & 85.08 (18.20) & 63.16 (60.21) & 50.13 (60.89) \\
  & & C & 55  & 110  & 4.25 (0.37) & 0.81 (0.39) & 0.89 (0.31) & 76.27 (19.01) & 83.86 (17.83) & 66.50 (58.55) & 57.99 (82.98) \\
\cmidrule(l){2-12}
  & \multirow{2}{*}{During (Q3--6)}
    & T & 60  & 240  & 4.37 (0.44) & 0.72 (0.45) & 0.82 (0.39) & 72.27 (22.17) & 83.77 (18.45) & 48.30 (43.73) & 22.45 (27.43) \\
  & & C & 55  & 220  & 4.25 (0.37) & 0.74 (0.44) & 0.85 (0.35) & 75.64 (21.43) & 80.89 (19.22) & 51.68 (46.07) & 42.78 (44.19) \\
\cmidrule(l){2-12}
  & \multirow{2}{*}{Eval (Q7--8)}
    & T & 60  & 120  & 4.37 (0.44) & 0.71 (0.46) & 0.87 (0.34) & 68.96 (26.15) & 76.46 (22.82) & 50.75 (42.96) & 39.90 (30.00) \\
  & & C & 55  & 110  & 4.25 (0.37) & 0.65 (0.48) & 0.78 (0.41) & 69.86 (25.57) & 77.73 (20.01) & 48.15 (54.36) & 43.04 (33.52) \\

\midrule

% NON-EXPERTS
\multirow{6}{*}{\textbf{Non-Experts}}
  & \multirow{2}{*}{Pre (Q1--2)}
    & T & 182 & 364  & 2.77 (0.70) & 0.74 (0.44) & 0.93 (0.26) & 56.63 (27.10) & 75.67 (22.32) & 62.31 (57.40) & 49.28 (42.20) \\
  & & C & 158 & 316  & 2.80 (0.69) & 0.74 (0.44) & 0.91 (0.28) & 55.35 (26.42) & 77.12 (20.58) & 59.50 (67.26) & 47.93 (44.93) \\
\cmidrule(l){2-12}
  & \multirow{2}{*}{During (Q3--6)}
    & T & 182 & 728  & 2.77 (0.70) & 0.74 (0.44) & 0.84 (0.37) & 50.84 (27.40) & 73.27 (21.79) & 53.14 (45.71) & 23.37 (30.47) \\
  & & C & 158 & 632  & 2.80 (0.69) & 0.71 (0.45) & 0.93 (0.26) & 54.07 (27.91) & 74.11 (21.47) & 51.61 (48.40) & 43.13 (44.72) \\
\cmidrule(l){2-12}
  & \multirow{2}{*}{Eval (Q7--8)}
    & T & 182 & 364  & 2.77 (0.70) & 0.60 (0.49) & 0.85 (0.36) & 46.80 (30.11) & 65.36 (26.29) & 48.19 (43.70) & 44.14 (33.75) \\
  & & C & 158 & 316  & 2.80 (0.69) & 0.61 (0.49) & 0.90 (0.31) & 47.28 (31.12) & 69.07 (24.30) & 47.14 (45.29) & 50.04 (48.08) \\

\bottomrule
\end{tabular}%
}%
{\NOTE{Notes.} Values are mean (SD). $N$ = number of participants; $Q$ = number of question-level observations;
T = AI abstention (treatment); C = Always Advising (control);
Exp.\ = perceived financial expertise (1--5 scale);
Init.\ Acc.\ = proportion of initial answers correct (pre-advice);
Final Acc.\ = proportion of final answers correct (post-advice);
Init.\ Conf.\ = initial confidence (0--100);
AI Conf.\ = confidence in AI advice (0--100);
Init.\ Time = initial response time in seconds (pre-advice);
Final Time = final response time in seconds (post-advice).
Experts = top 30\% of perceived expertise (mean $\geq$ 3.667, $N$ = 115).
Treatment $N$ = 242; Control $N$ = 213.}
\end{rotate}
\end{table}

Participants were recruited via Prolific using a stratified sampling approach designed to ensure adequate representation of users across the full range of perceived financial expertise. All participants were required to be based in the United States, fluent in English, aged 21 or older, and to have a Prolific approval rate of 95\% or above.

We recruited from three separate Prolific samples. The first was a general population sample of 400 participants with no additional screening criteria beyond the requirements above. Because general population samples tend to yield relatively few participants at the extremes of the expertise distribution, we supplemented this with two targeted samples of 100 participants each. The high-expertise sample required participants to hold an active trading platform account and to be either the primary financial decision maker in their household or to share equally in financial decision making. The low-expertise sample required participants to report having little to no involvement in household financial decisions.

This stratified design was intended to increase statistical power for detecting moderation effects at the extremes of the perceived expertise distribution, where treatment effects are theorized to be most pronounced. As a consequence, the analytic sample is not representative of the Prolific population or the general US population with respect to financial expertise. The oversampling of high- and low-expertise participants means that the marginal effects of AI abstention on acceptance reported in the main text should be interpreted as conditional on the observed expertise distribution in this sample rather than as population-average estimates. The moderation findings which are central to this work and which concern how the effect of AI abstention differs across expertise levels rather than its average magnitude, are not affected by this sampling design.

Table~\ref{tab:descriptive} report full descriptive statistics broken down by expertise group, experimental phase, and condition.
The table distinguishes three phases of the experiment: the pre-treatment phase (Questions 1--2), in which all participants received appropriate advice regardless of condition; the during phase (Questions 3--6), in which the treatment manipulation was administered; and the evaluation phase (Questions 7--8), in which both conditions received recommendations for one of the superior choices and advice acceptance was measured.

\section{Financial Decision Scenarios}
\label{appendix:scenarios}

The experiment included 10 financial decision scenarios covering investment allocation, retirement planning, credit card selection, stock valuation, debt management, and estate planning. Each scenario presented participants with four to five choice options following a dual-optimal structure, in which two options are designated as superior based on established financial principles and the remaining options are clearly inferior. The 10 scenarios are listed below ($^*$ denotes superior choices). 

\begin{enumerate}
    \item \textbf{Risk-Free Yield Comparison (sc3).} You are considering the best way to allocate some of your funds over the next 5 years. You are presented with the following risk-free yield options. Which of the following options would you select?
    \begin{itemize}
        \item Invest at an APR of 5\%, compounded annually$.^*$
        \item Invest at an APR of 4.94\%, compounded semi-annually$.^*$
        \item Invest at an APR of 4.8\%, compounded monthly.
        \item Invest at an APR of 4.7\%, compounded semi-annually.
        \item Invest at an APR of 4.8\%, compounded quarterly.
    \end{itemize}
    \textit{Rationale: Options 1 and 2 yield the highest effective annual returns when compounded interest is calculated correctly. Options 3, 4, and 5 yield lower effective returns despite some having higher nominal compounding frequency.}

    \item \textbf{401(k) Allocation (sc9).} You are 30 years old with \$20,000 in a 401(k), currently parked in a default money-market fund. You plan to leave these funds untouched for 30 years. Which of the following investment options do you choose?
    \begin{itemize}
        \item Put everything into a single tech company's stock.
        \item Allocate 60\% to a broad U.S. stock index fund and 40\% to a bond fund$.^*$
        \item Move everything into a gold-focused ETF.
        \item Keep it all in a short-term bond fund for safety.
        \item Invest the entire \$20,000 in a target-date fund that automatically adjusts over time$.^*$
    \end{itemize}
    \textit{Rationale: Options 2 and 5 reflect well-established long-horizon retirement investing principles -- diversification, low fees, and appropriate risk for a 30-year horizon. The remaining options are either overly concentrated, overly conservative, or speculative for a long-term retirement account.}

    \item \textbf{Declining Interest Rate Strategy (sc12).} You have noticed that interest rates have been declining. You currently have a mortgage, some credit card debt, and some emergency savings. Which of the following options is the best choice for you?
    \begin{itemize}
        \item Max out credit cards since rates are cheap.
        \item Consolidate high-interest debts with a low-interest loan$.^*$
        \item Ignore refinancing and keep paying original loan terms.
        \item Invest emergency savings into bonds.
        \item Refinance your mortgage at a lower rate$.^*$
    \end{itemize}
    \textit{Rationale: Options 2 and 5 both take rational advantage of declining interest rates to reduce borrowing costs. The remaining options are either financially harmful or miss the opportunity entirely.}

    \item \textbf{Portfolio Optimization (sc7).} You have \$15,000 to invest across five assets with varying expected returns and risk levels. Which of the following assets or combinations should you invest in for optimal performance?
    \begin{itemize}
        \item Diversify by buying equal amounts of each asset.
        \item All into Asset A (10\% return, average variance)$.^*$
        \item All into Asset C (10\% return, high variance).
        \item Half into Asset B, half into Asset D.
        \item Half into Asset A, half into Asset B$.^*$
    \end{itemize}
    \textit{Rationale: Options 2 and 5 offer the best risk-adjusted returns. Option 2 maximizes return for average risk; Option 5 balances high return with lower overall variance. Options 3, 4, and 1 either take on unnecessary risk or sacrifice return.}

    \item \textbf{Credit Card Selection (sc8).} You are looking for a credit card as your main payment method. You pay your balance in full each month, prefer rewards, and occasionally travel. Which of the following credit cards would you sign up for?
    \begin{itemize}
        \item Annual fee \$95, APR 16.99\%, 2x travel/dining rewards, 0\% foreign fee, 40,000 pt bonus$.^*$
        \item Annual fee \$300, APR 24.99\%, 1\% back, 3\% foreign fee, 40,000 pt bonus.
        \item Annual fee \$0, APR 17.99\%, 3\% back on groceries, 3\% foreign fee, \$200 bonus$.^*$
        \item Annual fee \$0, APR 25.99\%, 5\% off one retailer, 3\% foreign fee, \$20 bonus.
        \item Annual fee \$120, APR 22.99\%, 1\% back, 3\% foreign fee, no bonus.
    \end{itemize}
    \textit{Rationale: Options 1 and 3 offer the best combination of rewards, low fees, and sign-up bonuses for the described user profile. Options 2, 4, and 5 offer poor value relative to their costs.}

    \item \textbf{Portfolio Hedging Strategy (sc15).} You have a mid-sized portfolio that has significantly increased in value. Your risk tolerance is somewhat conservative and you expect a short-term market downturn but do not want to sell assets to avoid taxes. Which strategy should you use?
    \begin{itemize}
        \item Writing Naked Calls.
        \item Writing Covered Calls$.^*$
        \item Buying Long Calls.
        \item Iron Condor Spread.
        \item Buying Protective Puts$.^*$
    \end{itemize}
    \textit{Rationale: Options 2 and 5 are both appropriate hedging strategies for a conservative investor who wants downside protection without selling assets. Covered calls generate income; protective puts provide direct downside insurance.}

    \item \textbf{Stock Valuation (sc16).} You want to identify an undervalued mid-sized stock based on both fundamentals and growth potential. You have five candidates with varying P/E, DCF, liquidation value, and PEG ratios. Which stock is undervalued?
    \begin{itemize}
        \item Company A: P/E=38, DCF \$110 vs \$80, Liquidation \$50, PEG=0.8$.^*$
        \item Company B: P/E=10, DCF \$55 vs \$50, Liquidation \$30, PEG=1.6.
        \item Company C: P/E=15, DCF \$40 vs \$45, Liquidation \$25, PEG=1.1.
        \item Company D: P/E=60, DCF \$70 vs \$90, Liquidation \$40, PEG=2.5.
        \item Company E: P/E=14, DCF \$100 vs \$80, Liquidation \$60, PEG=0.9$.^*$
    \end{itemize}
    \textit{Rationale: Companies A and E show DCF fair value above current price and low PEG ratios, indicating undervaluation relative to growth. The other companies are either overvalued on DCF or have high PEG ratios indicating overpriced growth.}

    \item \textbf{Options Income Strategy (sc14).} You have a mid-sized portfolio and want to generate moderate income from options while limiting large losses. Your risk tolerance is moderate. Which strategy is reasonable?
    \begin{itemize}
        \item Writing Naked Calls.
        \item Writing Covered Calls$.^*$
        \item Buying Long Calls.
        \item Iron Condor Spread$.^*$
        \item Buying Protective Puts.
    \end{itemize}
    \textit{Rationale: Covered calls generate income with limited risk for a portfolio holder. The Iron Condor is a range-bound strategy that generates income with defined risk, appropriate for moderate risk tolerance.}

    \item \textbf{Estate Planning (sc13).} You are a 40-year-old single parent with a \$400,000 home and a \$100,000 retirement account. You want to name a guardian for your minor child, keep the process simple, and maintain control of assets during your lifetime. Which option fits your needs?
    \begin{itemize}
        \item Create an Irrevocable Trust.
        \item Create a Charitable Remainder Trust.
        \item Create a Testamentary Trust.
        \item Create a Simple Will$.^*$
        \item Create a Revocable Living Trust$.^*$
    \end{itemize}
    \textit{Rationale: A Simple Will and a Revocable Living Trust both allow guardian designation and straightforward asset distribution without complex tax strategies. The irrevocable and charitable options are unnecessarily complex for this estate size and profile.}

    \item \textbf{Investment Allocation (sc2).} You have \$5,000 to invest for at least 3--5 years. You are reasonably comfortable with market ups and downs but do not want extremely high volatility. How would you allocate the \$5,000?
    \begin{itemize}
        \item Invest in a single tech company's stock.
        \item Put it in a low-fee stock index fund$.^*$
        \item Keep the money in a high-yield savings account.
        \item Buy cryptocurrency.
        \item Split: large portion into a stock index fund, large portion into a bond fund, small portion into cryptocurrency$.^*$
    \end{itemize}
    \textit{Rationale: Options 2 and 5 offer appropriate diversification and risk levels for a moderate risk tolerance over a 3--5 year horizon. Options 1 and 4 are too speculative; option 3 is too conservative for the stated risk tolerance and time horizon.}
\end{enumerate}

\section{AI Response Generation and Examples}
\label{appendix:ai}

\subsection*{Generation Protocol}

AI explanations were developed through an iterative manual process using Claude Sonnet 4.5. We began by crafting an initial set of explanations and refining them until we were satisfied with their quality, accuracy, and consistency. We then used the established examples as templates to guide generation of the remaining responses, ensuring structural and tonal consistency across all scenarios. All explanations were manually reviewed for financial accuracy and cross-validated against established financial principles to confirm that optimal and suboptimal designations were correctly reflected in each response.

All responses share a consistent structure. Rather than opening with explicit agreement or disagreement, responses dive directly into the financial analysis. Agreeing responses open with a positive framing of the participant's choice and proceed to explain the financial principles that support it, with specific reference to the scenario constraints such as the participant's stated risk tolerance, time horizon, or financial goals. Disagreeing responses open by identifying the specific weakness of the chosen option, explain the financial reasoning that makes it suboptimal, and close by recommending a specific alternative option with a concise rationale for why it is superior. All responses are written in second person, reference the specific details of the scenario, and avoid generic advice. Responses are substantive but focused, typically 150--250 words. When advice was withheld, participants saw the message: ``No advice available.''

\subsection*{Example Responses}

The following examples illustrate the response format for Scenario 2 (Investment Allocation: \$5,000 over 3--5 years with moderate risk tolerance).

\textbf{Example agreeing response} (participant selected an optimal option -- low-fee stock index fund):

\begin{quote}
Investing in a low-fee stock index fund is an excellent strategy for your situation. With a 3--5 year time horizon, you have enough time to weather market fluctuations while benefiting from equity market growth. Index funds provide instant diversification across hundreds or thousands of companies, eliminating the concentration risk of individual stocks. The low fees mean more of your returns stay in your pocket rather than going to fund managers. This approach aligns perfectly with your comfort level --- you will experience normal market volatility but avoid the extreme swings associated with concentrated positions or speculative assets. Historically, broad market index funds have delivered solid returns over 3--5 year periods while maintaining reasonable risk levels. The simplicity and tax efficiency of this approach also work in your favor, and you are avoiding the behavioral pitfalls that come with trying to pick individual winners or time the market.
\end{quote}

\textbf{Example disagreeing response} (participant selected a suboptimal option -- single tech stock):

\begin{quote}
While investing in a single tech company's stock may offer high growth potential, this approach concentrates all your capital in one company, exposing you to company-specific risks like management decisions, competitive pressures, or sector downturns. With a 3--5 year time horizon, a single stock's volatility could be particularly pronounced, and you explicitly stated you do not want extremely high volatility. Concentration risk violates fundamental diversification principles --- if that one company underperforms, your entire investment suffers. Studies show that individual stock returns are far more volatile than diversified portfolios, and most individual stocks underperform broad market indices over time. Instead, I think a better approach would be putting your money in a low-fee stock index fund, which offers diversified exposure across hundreds of companies, reducing company-specific risk while still providing growth potential that aligns with your moderate risk tolerance and investment timeframe.
\end{quote}

\section{Condition Instructions and Comprehension Check}
\label{appendix:instructions}

\subsection*{Control Condition Instructions}

Participants assigned to the control condition received the following instructions prior to beginning the decision tasks:

\begin{quote}
In this study, you will see a series of personal finance scenarios. Each scenario will be followed by 4--5 possible choices. To assist in your decisions and support your performance, you will have access to an \textbf{AI financial advising model}. After your initial choice, you will see the AI's response and then submit your final answer.

\medskip
\noindent For each scenario, please: Read the scenario carefully. Choose the option you believe is best. See the AI's response. Submit your final answer --- you may update your choice or keep it.

\medskip
\noindent The AI advisor uses a comprehensive advising approach --- offering recommendations on every decision. Such AI advising systems are designed to provide their best advice across all situations. This approach has been shown to: provide broad coverage across all decisions; produce better downstream decisions; allow decision makers to evaluate and weigh AI recommendations for every case. When the AI provides a recommendation, it includes valuable information that can improve decision-making.
\end{quote}

\subsection*{Treatment Condition Instructions}

Participants assigned to the treatment condition received the following instructions:

\begin{quote}
In this study, you will see a series of personal finance scenarios. Each scenario will be followed by 4--5 possible choices. To assist in your decisions and support your performance, you will have access to an \textbf{AI financial advising model}. After your initial choice, you will see the AI's response and then submit your final answer.

\medskip
\noindent For each scenario, please: Read the scenario carefully. Choose the option you believe is best. See the AI's response. Submit your final answer --- you may update your choice or keep it.

\medskip
\noindent The AI advisor uses a selective advising approach --- offering recommendations only when they would be highly valuable to you. Selective AI advising systems are designed to recognize when algorithmic input is most likely to improve outcomes. This approach has been shown to: improve the quality of advice offered; produce better downstream decisions; allow decision makers to focus their attention on recommendations that are most beneficial. When the AI provides a recommendation, it reflects high-confidence, high-value input. When the AI does not provide advice, you will see: ``No advice available.''
\end{quote}

\subsection*{Comprehension Check}

Following the instructions, all participants completed a single comprehension check item confirming their understanding of the advising approach they would encounter. The item and correct responses for each condition are shown below.

\medskip
\noindent \textit{The AI in this study is designed to:}

\medskip
\noindent \textbf{Control condition} (correct answer in bold):
\begin{itemize}
    \item \textbf{Provide advice for every question.}
    \item Provide advice only when you request it.
    \item Never provide advice.
\end{itemize}

\medskip
\noindent \textbf{Treatment condition} (correct answer in bold):
\begin{itemize}
    \item \textbf{Provide advice selectively, only when its algorithmic input is most likely to improve outcomes.}
    \item Provide advice for every question.
    \item Provide advice only when you request it.
\end{itemize}

\noindent Participants who selected an incorrect response were excluded from the analytic sample.

\section{Survey Scales and Items}
\label{appendix:scales}

\subsection*{Pre-Survey Measures}

\subsection{Single-Item Perceived Financial Expertise}

Participants first responded to a single-item measure of perceived financial expertise used for treatment stratification:

\begin{quote}
\textit{How would you describe your overall knowledge and understanding of personal finance (e.g., budgeting, investing, saving, taxes, etc.)?}
\end{quote}

\noindent Response options: (1) I have very little knowledge of these topics; (2) I know some basics but still have a lot to learn; (3) I have a moderate understanding of most common financial topics; (4) I am very knowledgeable and regularly engage with financial topics; (5) I consider myself an expert in personal finance.

\subsection{Financial Expertise Scale}

Participants then completed three items adapted from the Consumer Financial Protection Bureau's Financial Skill Scale \citep{CFPB_FinancialSkillScale}. Participants rated how well each statement described them on a 5-point scale from (1) Not at all to (5) Completely.

\begin{enumerate}
    \item I know how to make complex financial decisions.
    \item I am able to recognize a good financial investment.
    \item I find it easy to understand financial information.
\end{enumerate}

\noindent The perceived financial expertise score used in all analyses is the mean of these three items ($\alpha = 0.897$, $N = 455$).

\subsection*{Post-Survey Measures}

\subsection{Overall Confidence}

Following all decision tasks, participants rated their overall confidence in two single items on a 5-point scale from (1) Strongly disagree to (5) Strongly agree:

\begin{enumerate}
    \item I have great confidence in the overall performance of the AI.
    \item I have great confidence in the overall accuracy of my answers.
\end{enumerate}

\subsection{Perceived Usefulness}

Three items measured perceived usefulness of the AI system on a 5-point scale from (1) Strongly disagree to (5) Strongly agree, adapted from \citet{Davis1989_MISQ_PerceivedUsefulness}:

\begin{enumerate}
    \item The AI reduced the effort required for making the decisions.
    \item The AI helped me improve my decision-making.
    \item Overall, I found the AI to be useful for making my decisions.
\end{enumerate}

\noindent ($\alpha = 0.838$, $N = 455$).

\subsection{Cognitive Burden}

Three items measured perceived cognitive burden of processing the AI's advice on a 5-point scale from (1) Strongly disagree to (5) Strongly agree. The items are adapted from \citet{KlepschSchmitzSeufert2017_FrontPsych_CognitiveLoadInstruments} and reworded to target the demand of evaluating and reconciling AI recommendations specifically:

\begin{enumerate}
    \item Understanding the AI was mentally demanding.
    \item It was difficult to assess when to rely on the AI.
    \item It took effort to recognize which information offered by the AI was important.
\end{enumerate}

\noindent The cognitive burden score used in all analyses is the mean of these three items ($\alpha = 0.773$, $N = 455$).

\subsection*{Additional Post-Survey Measures}

The following measures were collected for exploratory purposes and are not part of the primary analyses reported in the main text.

\subsection{Need for Cognition}

Three items measured dispositional need for cognition on a 5-point scale from (1) Strongly disagree to (5) Strongly agree ($\alpha = 0.880$, $N = 455$):

\begin{enumerate}
    \item I would prefer complex to simple problems.
    \item I like to have the responsibility of handling a situation that requires a lot of thinking.
    \item I really enjoy a task that involves coming up with new solutions to problems.
\end{enumerate}

\subsection{AI Sentiment}

Four items measured general attitudes toward AI on a 5-point scale from (1) Strongly disagree to (5) Strongly agree ($\alpha = 0.926$, $N = 455$):

\begin{enumerate}
    \item I believe that AI will improve my life.
    \item I believe that AI will improve my work.
    \item I think I will use AI technology in the future.
    \item I think AI technology is positive for humanity.
\end{enumerate}

\subsection{AI Engagement}

One item measured prior engagement with AI tools:

\begin{quote}
\textit{How much (if at all) have you previously engaged with AI tools (such as ChatGPT, Bing Copilot, Google's Gemini, Claude, etc.)?}
\end{quote}

\noindent Response options: (1) Not at all; (2) A little; (3) A moderate amount; (4) A lot; (5) A great deal.

\section{Pre-registration}
This study was preregistered on \href{ https://aspredicted.org/s5ib5i.pdf}{AsPredicted}.

\end{APPENDICES}

\end{document}